\documentclass[aps,prd,preprint,eqsecnum,tightenlines,floats,
showpacs,superscriptaddress]{revtex4}
\usepackage{amssymb} 
\usepackage{hyperref} 
\usepackage{graphicx}
\usepackage{subfigure}

\begin{document}


\date{November 4, 2011}

\title{Cosmic axion thermalization}

\author{O. Erken, P. Sikivie, H. Tam and Q. Yang}
\affiliation{Department of Physics, University of Florida, 
Gainesville, FL 32611, USA}

\begin{abstract}

Axions differ from the other cold dark matter candidates in that they 
form a degenerate Bose gas.  It is shown that their huge quantum degeneracy 
and large correlation length cause cold dark matter axions to thermalize 
through gravitational self-interactions when the photon temperature reaches 
approximately 500 eV.  When they thermalize, the axions form a Bose-Einstein 
condensate.  Their thermalization occurs in a regime, herein called the 
`condensed regime', where the Boltzmann equation is not valid because the 
energy dispersion of the particles is smaller than their interaction rate.  
We derive analytical expressions for the thermalization rate of particles 
in the condensed regime, and check the validity of these expressions by 
numerical simulation of a toy model.  We revisit axion cosmology in light 
of axion Bose-Einstein condensation.  It is shown that axions are 
indistinguishable from ordinary cold dark matter on all scales of 
observational interest, except when they thermalize or rethermalize.
The rethermalization of axions that are about to fall in a galactic 
potential well causes them to acquire net overall rotation as they go 
to the lowest energy state consistent with the total angular momentum 
they acquired by tidal torquing.  This phenomenon explains the occurrence 
of caustic rings of dark matter in galactic halos.  We find that photons 
may reach thermal contact with axions and investigate the implications 
of this possibility for the measurements of cosmological parameters.

\end{abstract}

\maketitle

\section{Introduction}

One of the outstanding problems in science today is the identity of the
dark matter of the universe \cite{PDM}.  The existence of dark matter is
implied by a large number of observations, including the dynamics of galaxy 
clusters, the rotation curves of individual galaxies, the abundances of light   
elements, gravitational lensing, and the anisotropies of the cosmic microwave 
background radiation.  The energy density fraction of the universe in dark
matter is observed to be 23\%.  The dark matter must be non-baryonic, cold 
and collisionless.  {\it Non-baryonic} means that the dark matter is not made 
of ordinary atoms and molecules.  {\it Cold} means that the primordial velocity
dispersion of the dark matter particles is sufficiently small, less than about 
$10^{-8}~c$ today, so that it may be set equal to zero as far as the formation 
of large scale structure and galactic halos is concerned.  {\it Collisionless} 
means that the dark matter particles have, in first approximation, only 
gravitational interactions.  Particles with the required properties are 
referred to as `cold dark matter' (CDM).  The leading CDM candidates are 
weakly interacting massive particles (WIMPs) with mass in the 100 GeV range, 
axions with mass in the $10^{-5}$ eV range, and sterile neutrinos with mass 
in the keV range.  To  try and tell these candidates apart on the basis of 
observation is a tantalizing quest.

For a long time, it was thought that axions and the other forms of cold 
dark matter behave in the same way on astronomical scales and are therefore 
indistinguishable by observation, whether it be observations of large scale 
structure or measurements of cosmological parameters.  More recently, however,
it was  pointed out that dark matter axions form a Bose-Einstein condensate 
(BEC), as a result of their gravitational self-interactions, when the photon 
temperature reaches about 500 eV \cite{CABEC}.  Axions or axion-like particles 
are special because they are a degenerate Bose gas.  The other dark matter 
candidates, which we refer to henceforth as `ordinary cold dark matter', are 
non-degenerate.  This raises the question whether axions may be observably 
different after all.  It was shown in ref. \cite{CABEC} that, on all scales
of observational interest, density perturbations in axion BEC behave in 
exactly the same way as those in ordinary cold dark matter provided the
density perturbations are within the horizon and in the linear regime.  On 
the other hand, when density perturbations enter the horizon, or in second 
order of perturbation theory, axions generally behave differently from 
ordinary cold dark matter because the axions rethermalize to let the axion 
state (i.e. the state most axions are in) track the lowest energy state 
\cite{CABEC}.  

Axion Bose-Einstein condensation appears to resolve a puzzle that has arisen 
in the study of the inner caustics of galactic halos.  The structure of the 
inner caustics depends on the angular momentum distribution of the infalling
particles.  If the particles fall in with net overall rotation, the inner 
caustics are rings whose cross-section is a section of the elliptic umbilic 
($D_{-4}$) catastrophe, called `caustic rings' for short \cite{crdm,sing}.  
If the velocity field of the infalling particles is irrotational, the inner 
caustics have a `tent-like' structure which is described in detail in ref. 
\cite{inner} and which is quite distinct from that of caustic rings.  The 
radii of the caustic rings, assuming that the dark matter falls in with net 
overall rotation, were predicted \cite{crdm} using the self-similar infall 
model of galactic halos \cite{FGB}, generalized to allow angular momentum 
for the infalling particles \cite{STW,MWhalo}.  Evidence was found for the 
existence of caustic rings at the predicted radii.  The evidence is summarized 
in ref. \cite{MWhalo}.  Now, the puzzle is that ordinary cold dark matter has 
an irrotational velocity field \cite{inner} and is therefore incompatible with 
the existence of caustic rings.  Axion Bose-Einstein condensation resolves the 
puzzle provided the axions rethermalize sufficiently quickly that most of them 
go to the lowest energy available state before falling in \cite{CABEC,case}.  
The lowest energy state consistent with the total angular momentum the axions 
will have acquired from neighboring inhomogeneities by tidal torquing is a 
state of rigid rotation on the turnaround sphere, the simplest form of net 
overall rotation.  Therefore, if the dark matter is an axion BEC that 
rethermalizes sufficiently quickly, the inner caustics are rings.  Furthermore 
it was shown \cite{case} that the caustic rings are all in the same plane 
(that of the galactic disk), that the overall size of the rings is predicted 
correctly by tidal torque theory, and that the relative sizes of the rings 
are precisely as predicted by the self-similar model and therefore consistent 
with the evidence for caustic rings published earlier.   One thing remains 
to be done: to show that axions about to fall into a galactic potential well 
rethermalize sufficiently quickly that they almost all go to the lowest energy 
available state.  It is one of the goals of this paper to understand cosmic 
axion thermalization sufficiently well to be able to verify this.

There is a second motivation for studying the thermalization rates of dark 
matter axions in detail\cite{Li7}.  We mentioned that cold axions thermalize 
by their gravitational interactions.  But gravity is universal.  If axions 
thermalize by gravitational interactions, they may also enter into thermal 
contact with other species.  If the axions enter into thermal contact with 
the cosmic photons, the photons will cool and some cosmological parameters, 
in particular the baryon to photon ratio at primordial nucleosynthesis and 
the effective number of neutrinos (a measure of the radiation density at the 
time of decoupling), will be modified compared to their values in the standard
cosmological model.  This opens the possibility of being able to distinguish 
axions from the other forms of cold dark matter by measuring cosmological 
parameters.

Due to their unusual properties, dark matter axions occupy a unique and 
uncharted region in the physics of many-body systems. On the one hand, 
they are highly condensed, which greatly exaggerates the quantum effect of 
Bose-enhancement in scattering processes. On the other, the fact that their 
energy dispersion is very small implies that they are outside the realm of 
the `particle kinetic regime'. In this case, the picture of instantaneous 
collisions breaks down, so that the usual Boltzmann equation no longer 
applies. Together, these properties -- high occupation number and small 
energy dispersion -- are highly atypical, and thus very little attention 
has been paid to the study of such systems.  Existing techniques in 
non-equilibrium statistical mechanics are not applicable to dark 
matter axions, rendering the estimation of their thermalization 
rate non-trivial.

The outline of this paper is as follows.  In Section II, we give a 
definition of cold dark matter and ask under what conditions axions 
behave as such.  We find that axions behave as CDM on all scales of 
observational interest except when they thermalize or rethermalize.
In Section III, we derive expressions for the thermalization rate of 
particles in the `condensed regime', i.e. when their energy dispersion 
is small compared to their interaction rate.  The condensed regime is
the one relevant to cosmic axion thermalization.  In Section IV, we 
check the validity of our estimates of thermalization rates in the 
condensed regime by numerical simulation of a toy model.  In Section V, 
we revisit axion cosmology in light of axion thermalization and 
Bose-Einstein condensation.  Section VI provides a summary.

\section{Axions are different}

Both axions and WIMPs are considered forms of cold dark matter.  
Furthermore, until recently, axions and WIMPs were thought to be 
indistinguishable on observational grounds, i.e. indistinguishable 
on the basis of purely astronomical data.  The discovery \cite{CABEC} 
that dark matter axions form a Bose-Einstein condensate has changed 
this view since axion BEC is claimed to have observable consequences
\cite{CABEC,case,Li7}.  This raises questions.  First, precisely under 
what conditions are axions and WIMPs cold dark matter?  Second, if both 
axions and WIMPs are CDM, how do they differ?  There is a preliminary 
question: how is cold dark matter defined?

The purpose of this section is to discuss these three questions in a 
general way, to set the stage for the more detailed calculations that 
follow. For the sake of brevity, we call WIMPs all cold dark matter 
candidates which do not form a degenerate Bose gas.  As far as we are 
aware, this includes all cold dark matter candidates except axions 
and axion-like particles.  In particular, {\it sterile} neutrinos 
are called WIMPs here. (Neutrinos are not WIMPs since they are hot 
dark matter.)

To start off it is worth emphasizing that, at the fundamental level, 
axions and WIMPs are very different.  The surprise is really that they
have similar properties as far as large scale structure is concerned.
Both axions and WIMPs are described by quantum fields.  Furthermore, 
both are excellently described by classical limits of quantum fields.  
But the classical limits are different in the two cases:  WIMPs are in 
the classical particle limit whereas (decoupled) axions are in the classical 
field limit.  In the classical particle limit one takes $\hbar \rightarrow 0$ 
while keeping $E = \hbar \omega$ and $\vec{p} = \hbar \vec{k}$ fixed.  Since 
$\omega, \vec{k} \rightarrow \infty$, the wave nature of the quanta disappears.  
WIMPs are to excellent approximation classical point particles.  In the 
classical field limit, on the other hand, one takes $\hbar \rightarrow 0$ 
for constant $\omega$ and $\vec{k}$.  $E = {\cal N} \hbar \omega$ and 
$\vec{p} = {\cal N} \hbar \vec{k}$ are held fixed by letting the quantum 
state occupation number ${\cal N} \rightarrow \infty$.  This is the limit 
in which quantum electrodynamics becomes classical electrodynamics.  It is 
the appropriate limit for (decoupled) cold dark matter axions because they 
are a highly degenerate Bose gas.  The axion states that are occupied have 
huge occupation numbers, ${\cal N} \sim 10^{61}$ \cite{CABEC}.  The need 
to restrict to {\it decoupled} axions will be explained shortly.

So axions and WIMPs are fundamentally different even if it turns out 
that both can legitimately be called CDM.  The distinction is not just 
academic, and is certainly important if axions thermalize, i.e. if axions 
find a state of larger entropy through self-interactions.  Recall that, 
whereas statistical mechanics makes sense of the behaviour of large 
aggregates of classical particles (it was invented by Boltzmann to derive 
the properties of atoms in the gaseous state) it fails to make sense of 
classical fields. In thermal equilibrium every mode of a classical 
field would have average energy $k_{\rm B} T$.  As Rayleigh pointed out, 
the energy density is infinite then at finite temperature due to the 
contributions from short wavelength modes.  Thus the application of 
statistical mechanics to classical field theory (classical electrodynamics 
in particular) is in direct disagreement with observation.  As is well-known, 
the disagreement is removed because of, and only because of, quantum mechanics.   

What do we learn from this?  If the axions are decoupled (i.e. do not interact 
and hence do not thermalize), they behave to excellent approximation like 
classical fields.  As such, they are quite different from WIMPs since WIMPs 
are classical particles.  So the argument why decoupled axions behave as CDM 
must be different from the argument why WIMPs behave as CDM.  If the axions 
thermalize, they are not even described by classical fields.  Instead they 
form a Bose-Einstein condensate, an essentially quantum-mechanical phenomenon.  
If the axions form a BEC, the argument why the axion BEC behaves as CDM must 
be different again.

The main purpose of this paper is, in fact, to show that axions thermalize 
as a result of their gravitational interactions, that they form a BEC, and 
that a rethermalizing axion BEC behaves differently from CDM.  This will be 
done in Sections III, IV and V.  In the remainder of the present section, 
we first adopt a definition of cold dark matter, next we discuss under what 
conditions WIMPs behave as CDM, and then discuss under what conditions 
axions behave as CDM.

\subsection{Cold dark matter}

Astronomical data, in particular data on the cosmic microwave 
background anisotropies and on the large scale structure of 
the universe in the linear regime, imply the existence of a  
new kind of stuff called `cold dark matter' \cite{PDM}.  What 
explains the data may technically be called a {\it perfect 
fluid with zero pressure} \cite{Dodel}.   The state of such 
a fluid is characterized by a density $\rho(\vec{r}, t)$ and 
a velocity field $\vec{v}(\vec{r}, t)$.  These four degrees 
of freedom satisfy the continuity equation
\begin{equation}
\partial_t \rho + \vec{\nabla} \cdot (\rho~\vec{v}) = 0 
\label{coneq}
\end{equation} 
and, in the Newtonian limit of gravity, the equation of motion
\begin{equation}
\partial_t \vec{v} + (\vec{v} \cdot \vec{\nabla}) \vec{v} 
= - \vec{\nabla} \Phi~~~~\ .
\label{eom}
\end{equation}
$\Phi$ is the gravitational potential.  All forces on the 
CDM fluid other than gravity are assumed to be negligible. 
The CDM fluid is a source for the gravitational potential:
\begin{equation}
\nabla^2~\Phi(\vec{r}, t) = 4 \pi G (\rho(\vec{r}, t) + ... )
\label{Laplace}
\end{equation}
where the dots represent other sources.  The stress-energy-momentum 
is given by 
\begin{equation}
{\cal T}^{00} = \rho~~~~,~~~~{\cal T}^{0k} = \rho v^k~~~~,
~~~~{\cal T}^{kl} = \rho v^k v^l~~~\ .
\label{sem}
\end{equation}
We will take the above to be the defining properties of CDM because 
stuff with those properties explains the data mentioned \cite{Dodel}.  
We will not concern ourselves with the very interesting question to 
what extent the data demand CDM to be pressureless and collisionless.  
Instead we want to ask in what limit WIMPs and axions have the 
properties stated above.  

To show that WIMPs or axions behave as CDM it is sufficient to show that, 
in some average sense, they have the same stress-energy-momentum tensor, 
Eq. (\ref{sem}).  Indeed the conservation law $D_\mu {\cal T}^{\mu\nu} = 0$ 
is equivalent in the Newtonian limit to Eqs.~(\ref{coneq}) and (\ref{eom}) 
whereas Einstein's equations imply Eq.~(\ref{Laplace}).

For the sake of clarity, let us emphasize that the above is only a 
good definition of CDM in the linear regime of structure formation.  
In the non-linear regime CDM produces discrete flows \cite{Ips,rob} 
and caustics \cite{crdm,sing} and these cannot be described by Eqs. 
(\ref{coneq} - \ref{Laplace}).  Instead the state of CDM in the 
non-linear regime is described by the embedding of a 3-dim. hypersurface 
in 6-dim. phase space plus the density of particles on this 3-dim. 
hypersurface.  This more general (and hence better) description of 
CDM defines it as a {\it cold collisionless fluid}.  In the linear 
regime, the two descriptions are equivalent because the 3-dim. 
hypersurface covers physical space only once.  

\subsection{WIMPs}

WIMP dark matter is a (huge) collection of classical point particles. 
Let their number be $N$.  Their state is given at time $t$ by giving 
all the particle positions $\vec{r}_i(t)$ and velocities $\vec{v}_i(t)$, 
$i = 1, 2, ... ~N$.  Their time evolution from appropriate initial 
conditions is determined by the equations of motion:
\begin{equation}
{d^2 \vec{r}_i \over dt^2} = {d \vec{v}_i \over dt} 
= - \vec{\nabla} \Phi(\vec{r}_i, t)~~~\ .
\label{Weom}
\end{equation}
The stress-energy-momentum tensor is
\begin{equation}
(T^{00}~,~T^{0k}~,~T^{kl})(\vec{r}, t) 
= m_W \sum_{j=1}^N (1~,~v_j^k(t)~,~v_j^k(t) v_j^l(t)) 
\delta(\vec{r} - \vec{r}_j(t)) ~~~\ ,
\label{WT}
\end{equation}
where $m_W$ is the WIMP mass.  The indices $i$ and $j$ label the particles 
whereas $k,l = 1, 2, 3$ label the spatial directions.  We want to derive the
conditions under which the WIMP particles behave as CDM, defined in the previous 
subsection.  One goes from the one description to the other by averaging over 
a volume $V$ centered at $\vec{r}$:
\begin{eqnarray}
\rho(\vec{r}, t) &\equiv& {\cal T}^{00} \equiv 
{1 \over V} \int_V d^3 s~T^{00}(\vec{r} + \vec{s}, t) 
= {m_W \over V} \int_V d^3 s \sum_{j=1}^N 
\delta(\vec{r} + \vec{s} - \vec{r}_j(t))\nonumber\\
\vec{\cal P}(\vec{r}, t) &\equiv& \hat{k} {\cal T}^{0k} \equiv 
{1 \over V} \int_V d^3 s~\hat{k}~T^{0k}(\vec{r} + \vec{s}, t) 
= {m_W \over V} \int_V d^3 s \sum_{j=1}^N~\vec{v}_j(t)~ 
\delta(\vec{r} + \vec{s} - \vec{r}_j(t))\nonumber\\
{\cal T}^{kl}(\vec{r}, t) &\equiv& {1 \over V}
 \int_V d^3 s~T^{kl}(\vec{r} + \vec{s}, t)
= {m_W \over V} \int_V d^3 s \sum_{j=1}^N~v_j^k(t)~v_j^l(t)~
\delta(\vec{r} + \vec{s} - \vec{r}_j(t))~~~~\ .
\label{Wav}
\end{eqnarray}
We take $V$ to be time independent.  $V$ must be large enough 
that the fluctuations in particle number inside $V$ are negligible.
Conservation of the averaged stress-energy-momentum ${\cal T}^{\mu\nu}$ 
follows merely from the conservation of the original stress-energy-momentum 
tensor $T^{\mu\nu}$.  But ${\cal T}^{\mu\nu}$ does not generally have the 
perfect fluid form.  The velocity of the WIMP fluid is 
\begin{equation}
\vec{v}(\vec{r}, t) = 
{1 \over \rho(\vec{r}, t)} \vec{\cal P}(\vec{r}, t)~~\ .
\label{Wvel}
\end{equation}
For each particle inside volume $V$, let us define
\begin{equation}
\Delta \vec{v}_i (\vec{r}, t) \equiv  \vec{v}_i(t) - \vec{v}(\vec{r}, t)~~~\ .
\label{deltavel}
\end{equation}
We have then 
\begin{equation}
{\cal T}^{kl}(\vec{r}, t) = 
\rho(\vec{r}, t)~v^k(\vec{r}, t)~v^l(\vec{r}, t)
+ {m_W \over V} \int_V d^3 s \sum_{j=1}^N~\Delta v_j^k~\Delta v_j^l
\delta(\vec{r} + \vec{s} - \vec{r}_j(t))~~~~\ . 
\label{notquite}
\end{equation}
If the velocity distribution inside $V$ is isotropic, we 
may define the pressure $p(\vec{r}, t)$ by 
\begin{equation}
{m_W  \over V} \int_V d^3 s \sum_{j=1}^N~\Delta v_j^k~\Delta v_j^l                     
\delta(\vec{r} + \vec{s} - \vec{r}_j(t)) 
\equiv p(\vec{r}, t) \delta^{kl}
\label{pressure}
\end{equation}
so that 
\begin{equation}
{\cal T}^{kl} = \rho(\vec{r}, t)~v^k(\vec{r}, t)~v^l(\vec{r}, t)
+ p(\vec{r}, t) \delta^{kl}~~~\ .
\label{pff}
\end{equation}
The stress-energy-momentum tensor is then said to have the 
perfect fluid form.  If, in addition, there is a relation 
$p(\rho)$ that determines pressure in terms of density 
(e.g. the relation $p(\rho)$ for atomic gases implied 
by adiabatic expansion/compression), the conservation laws
$D_\mu T^{\mu\nu} = 0$, being four equations for the four 
unknowns $\rho(\vec{r}, t)$ and $\vec{v}(\vec{r}, t)$, allow 
us to determine the evolution from arbitrary initial conditions.

As mentioned, CDM is defined to be a {\it pressureless} perfect 
fluid.  The pressure is a measure of the velocity dispersion of 
the particles: $p = {\rho \over 3} <(\Delta \vec{v})^2>$.  For 
WIMPs to be CDM, their velocity dispersion must be sufficiently 
small.  The best limit comes from the fact that WIMP density 
perturbations on scales less than their free streaming distance 
are erased.  Too large a velocity dispersion is inconsistent 
with the existence of the smallest observed large scale structure, 
that which gives rise to the `Lyman-alpha forest'.  This constrains 
the primordial WIMP velocity dispersion to be less than of order 
$10^{-8}c$ today.  If the WIMP particles have ordinary weak 
interactions, their kinetic energy decouples in the early 
universe ({\it i.e.} they stop colliding with other particles
in the primordial soup) when the photon temperature is a few
MeV.  Their velocity dispersion today is then 
\begin{equation}
\sqrt{<(\Delta \vec{v})^2>} ~\sim~ 10^{-12}~c~
\left({100~{\rm GeV} \over m_W}\right)^{1 \over 2}~~~\ .
\label{Wvd}
\end{equation}
The lightest allowed mass is therefore of order keV, which is 
that of sterile neutrinos \cite{ster}.

In summary, WIMPs behave as CDM if there exists a volume $V$
large enough so that the fluctuations in particle number inside 
the volume are negligible, and small enough that the velocity 
dispersion of the particles inside the volume is negligible.
Under these conditions, the WIMPs behave as CDM for any observer
who does not resolve length scales smaller than $V^{1 \over 3}$.

Finally, let us remark that gravitational interactions among 
point particles is a source of velocity dispersion:
\begin{equation}
\Delta v|_g \sim \sqrt{G m_W \over d_W} =
\sqrt{G \rho}~d_W \sim ~ 10^{-26} c 
\left({m_W \over 100~{\rm GeV}}\right)^{1 \over 3}~~~\ ,
\label{gvd}
\end{equation}
where $d_W$ is the average interparticle distance.
It is not possible for the velocity dispersion to be 
less than $\Delta v|_g$.  Furthermore $\Delta v|_g$
grows by gravitational instability as the particles aggregate
into clumps of ever increasing size.  $\Delta v|_g$
is completely negligible for all proposed WIMP candidates,
such as neutralinos and sterile neutrinos.  However in 
present numerical simulations of structure formation the 
particle mass is $10^4$ solar masses or more and hence 
$\Delta v|_g$ is larger than $4 \cdot 10^{-7}$, which is 
more than the allowed velocity dispersion of CDM.  By this
criterion and others \cite{rob}, present simulations do 
not have sufficient resolution to describe CDM.  For the 
particles in the simulations to have velocity dispersion 
as low as that required of CDM, the particle mass should
be no more than approximately one solar mass.

\subsection{Axions}

In the first part of this subsection, we give the argument why 
{\it decoupled} cold axions behave as CDM.  Cold axions remain 
decoupled in the early universe from shortly after their first 
appearance during the QCD phase transition till the time when 
gravitational self-interactions cause them to form a BEC.  The 
Bose-Einstein condensation of axions occurs when the photon 
temperature is approximately 
$500~{\rm eV} \left(f_a \over {10^{12}~{\rm GeV}}\right)^{1 \over 2}$
(see Section V).  After the condensation is complete, almost all 
axions are in the same state.  In the second part of this subsection, 
we give the argument why axions behave as CDM when they are all in 
the same state and that state does not change by rethermalization.

The results obtained in this subsection are not new, although some 
of the derivations may be.  Descriptions of CDM in terms of a classical 
field are given in refs. \cite{axdm,axgal,Ruff,Hwang}.  Descriptions of 
CDM in terms of the Schr\"odinger equation are given in refs. \cite{WKB},
\cite{CABEC}. 

\subsubsection{Decoupled axions}

As mentioned earlier, decoupled axions behave as classical 
fields because the quantum state occupation numbers of those
states that are occupied are huge, of order $10^{61}$ \cite{CABEC}.  
In flat space-time, the classical axion field $\phi(\vec{r}, t)$ 
satisfies the equation of motion 
\begin{equation}
(\partial_t^2 - \vec{\nabla}^2 + m^2) \phi = 0
\label{weq}
\end{equation}
where $m$ is the axion mass.   Space-time curvature does 
not play a role in this discusssion, so we ignore it.  The 
stress-energy-momentum tensor is
\begin{eqnarray}
T^{00} &=& {1 \over 2}\left( \dot{\phi}^2 + (\vec{\nabla} \phi)^2 
+ m^2 \phi^2 \right)\nonumber\\
\hat{k} T^{0k} &=& - \dot{\phi} \vec{\nabla}\phi\nonumber\\
T^{kl} &=& \partial_k \phi~\partial_l \phi 
+ {1 \over 2} \left(\dot{\phi}^2 - (\vec{\nabla} \phi)^2
- m^2 \phi^2 \right) \delta_{kl}~~~\ .
\label{axT}
\end{eqnarray}
To find out under what conditions decoupled axions behave as CDM, 
we average again over a spatial volume $V$ located at $\vec{r}$:
\begin{equation}
{\cal T}^{\mu\nu}(\vec{r}, t)  \equiv 
{1 \over V} \int_V d^3 s~T^{\mu\nu} (\vec{r} + \vec{s}, t)~~~\ ,
\label{avaxT}
\end{equation} 
where $\mu, \nu = 0, 1, 2, 3$.  To perform the average, we Fourier 
transfom $\phi$ within the volume $V$:
\begin{equation}
\phi(\vec{x}, t) = \sum_{\vec{p}}~
\left(\phi(\vec{p}, \vec{r}, t) e^{ i p \cdot x} +
\phi^*(\vec{p}, \vec{r}, t) e^{ - i p \cdot x} \right)~~~\ ,
\label{Four}
\end{equation}
where $p \cdot x = - p^0 t + \vec{p} \cdot \vec{x}$ and $p^0 = 
+ \sqrt{(\vec{p})^2 + m^2}$.  The Fourier components depend 
on the position $\vec{r}$ of the volume.  One finds:
\begin{eqnarray}
\rho(\vec{r}, t) &\equiv& {\cal T}^{00} = 
\sum_{\vec{p}}~N(\vec{p}, \vec{r}, t) p^0\nonumber\\
\vec{\cal P}(\vec{r}, t) &\equiv& \hat{k} {\cal T}^{0k} = 
\sum_{\vec{p}}~N(\vec{p}, \vec{r}, t) \vec{p}\nonumber\\
{\cal T}^{kl} &=& \sum_{\vec{p}}~
N(\vec{p}, \vec{r}, t) p^k p^l {1 \over p^0}~~~\ ,
\label{aT}
\end{eqnarray}
where 
\begin{equation}
N(\vec{r}, \vec{p}, t) = 2 p^0 |\phi(\vec{p}, \vec{r}, t)|^2~~~\ .
\label{phspden}
\end{equation}
To obain the expression for ${\cal T}^{kl}$ it is necessary to 
average not only over the volume $V$, but also over time.  The 
time interval to be averaged over should be much longer than 
$m^{-1}$.  Eqs.~(\ref{aT}) show that 
\begin{equation}
{\cal N}(\vec{r}, \vec{p}, t) = 
{V \over (2 \pi)^3} N(\vec{r}, \vec{p}, t)
\label{curlyN}
\end{equation}
may be thought of as the phase space density of the coarse grained 
axion field.

We assume the axions to be non-relativistic.  The average
momentum at $\vec{r}$ is\\ $\langle\vec{p}\rangle(\vec{r}, t) =
{m \over \rho} \vec{\cal P}$.  The classical axion field behaves 
like CDM provided the momentum distribution $N(\vec{r}, \vec{p}, t)$ 
is narrowly peaked around $\langle\vec{p}\rangle(\vec{r}, t)$ for all 
$\vec{r}$.  The velocity field is then 
$\vec{v}(\vec{r}, t) = {1 \over m} \langle\vec{p}\rangle(\vec{r},t) = 
{1 \over \rho} \vec{\cal P}$ and 
\begin{equation}
{\cal T}^{kl} (\vec{r}, t) = 
\rho(\vec{r}, t) v^k(\vec{r}, t) v^l(\vec{r}, t) +
\sum_{\vec{p}} N(\vec{r}, \vec{p}, t) {1 \over m} 
\delta p_k(\vec{r}, \vec{p}, t) \delta p_l(\vec{r}, \vec{p}, t)
\label{fas}
\end{equation}
where 
\begin{equation}
\delta {\vec p}(\vec{r}, \vec{p}, t) = \vec{p} - 
\langle\vec{p}\rangle(\vec{r}, t)~~~\ .
\label{delp}
\end{equation}
Therefore one condition for decoupled axions to behave as CDM is that 
they have sufficiently small velocity dispersion.  That condition is 
satisfied for the axions produced during the QCD phase transition.
Indeed their velocity dispersion is [see Section V.A]
\begin{equation}
\delta v (t) \sim {1 \over m t_1} {a(t_1) \over a(t)}
\label{avd}
\end{equation}
where $t_1 \simeq 2 \cdot 10^{-7} {\rm sec} 
\left({f_a \over 10^{12}~{\rm GeV}}\right)^{1 \over 3}$ is the time 
when the axion mass effectively turns on during the QCD phase transition, 
and $a(t)$ is the cosmological scale factor.  If these axions remained 
decoupled afterward, their velocity dispersion today would be approximately 
$5 \cdot 10^{-17}~c~\left({f_a \over 10^{12}~{\rm GeV}}\right)^{5 \over 6}$, 
certainly small enough to be called CDM.

The second condition for decoupled axions to behave as CDM is that they 
be observed only on length scales larger than their correlation length.
The low velocity dispersion of cold decoupled axions implies a 
correspondingly large correlation length
\begin{equation}
\ell(t) = {1 \over m \delta v(t)} \sim t_1 {a(t) \over a(t_1)}~~~\ .
\label{corrl}
\end{equation}
On scales less than $\ell(t)$, axions behave differently from CDM. 
Indeed for CDM, $\rho(\vec{r}, t)$ and $\rho(\vec{r}~^\prime, t)$ are 
independent variables no matter how close $\vec{r}$ and $\vec{r}~^\prime$, 
whereas for decoupled axions, $\rho(\vec{r}, t)$ and $\rho(\vec{r}~^\prime, t)$ 
are independent variables only if $|\vec{r} - \vec{r}~^\prime| > \ell$.  
However, even today, $\ell$ is only about 
$10^{17} {\rm cm} \left({f_a \over 10^{12}~{\rm GeV}}\right)^{1 \over 6}$,
much smaller than any scale on which we have observational information on the 
nature of CDM.

\subsubsection{All axions in the same quantum state}

When all the axions are and remain in the same quantum state, the axion 
fluid also behaves as CDM on all scales of observational interest.  The 
argument for this was given in ref. \cite{CABEC}, but we restate it here 
for the sake of completeness.  Note there is no requirement that the axion 
state is the lowest energy state.  In principle the axion state may be any 
state.  It is important, however, that the axions remain in the same state 
all the time.

The quantum axion field may be expanded in modes labeled $\vec\alpha$:
\begin{equation} 
\phi(x) = \sum_{\vec\alpha}~[a_{\vec\alpha}~\Phi_{\vec\alpha}(x)  
~+~a_{\vec\alpha}^\dagger~\Phi_{\vec\alpha}^*(x)] 
\label{modex} 
\end{equation} 
where the $\Phi_{\vec\alpha}(x)$ are the positive frequency c-number 
solutions of the Heisenberg equation of motion for the axion field 
\begin{equation}
D^\mu D_\mu \varphi(x) = 
g^{\mu\nu}[\partial_\mu \partial_\nu - \Gamma_{\mu\nu}^\lambda
\partial_\lambda] \varphi(x) = m^2 \varphi(x)~~~\ , 
\label{feq} 
\end{equation} 
and the $a_{\vec\alpha}$ and $a_{\vec\alpha}^\dagger$ are annihilation 
and creation operators satisfying canonical commutation relations.  
Eqs.~(\ref{feq}) are written in curved space-time to emphasize that 
the modes depend on the background.  The words `mode' and `particle 
state' are equivalent.  Let us assume that, except for a tiny fraction, 
all axions go to a single particle state with label $\vec\alpha = 0$.  
The corresponding $\Phi_0(x)$ is the axion wavefunction.  For the sake 
of brevity, we call the state of the axion fluid with almost all axions 
in a single particle state an axion BEC whether or not that particle 
state is the lowest energy state.  

Ignoring the small fraction of axions that are not condensed into mode 
$\vec\alpha = 0$, the state of the axion field is 
$|N> = (1/\sqrt{N!})~(a_0^\dagger)^N |0>$ where $|0>$ is the 
empty state, defined by $a_{\vec\alpha}~|0>$ = 0 for all $\vec\alpha$, 
and $N$ is the number of axions.  The expectation value of the 
stress-energy-momentum tensor is  
\begin{equation}
{\cal T}_{\mu\nu} (\vec{r}, t) \equiv <N|T_{\mu\nu}|N> = N
[\partial_\mu \Phi_0^* \partial_\nu \Phi_0 
+ \partial_\nu \Phi_0^* \partial_\mu \Phi_0 +
g_{\mu\nu} ( - \partial_\lambda \Phi_0^* \partial^\lambda \Phi_0  
- m^2 \Phi_0^* \Phi_0)]~~~\ .
\label{Tmunu}
\end{equation}
To see under what conditions the axion BEC behaves as CDM, we may
restrict ourselves to Minkowski space-time.   Since the axions are 
non-relativistic, $\Phi_0(x) = e^{- i m t} \Psi(x)$ with $\Psi(x)$ 
slowly varying.  Neglecting terms of order ${1 \over m}~\partial_t$ 
compared to terms of order one, Eq.~(\ref{feq}) becomes the Schr\"{o}dinger 
equation:
\begin{equation}
i \partial_t \Psi = - {\nabla^2 \over 2 m} \Psi~~~~\ .
\label{Schrod}
\end{equation}
The wavefunction may be written as \cite{Pethik}
\begin{equation}
\Psi(\vec{x}, t) = {1 \over \sqrt{2 m N}} B(\vec{x}, t)
e^{i \beta(\vec{x}, t)}~~~\ .
\label{dec}  
\end{equation}
In terms of $B(\vec{x}, t)$ and $\beta(\vec{x}, t)$ 
the energy and momentum densities are 
${\cal T}_{00} \equiv \rho = m \left(B(\vec{x}, t)\right)^2$   
and ${\cal T}_{0j} \equiv - \rho v_j =
- \left(B(\vec{x}, t)\right)^2 \partial_j \beta$,
in the non-relativistic limit.  The velocity field is therefore
$\vec{v}(\vec{x}, t) = {1 \over m} \vec\nabla \beta(\vec{x}, t)$
\cite{Pethik}.  Eq.~(\ref{Schrod}) implies the continuity equation,
Eq.~(\ref{coneq}), and the equation of motion
\begin{equation}
\partial_t v^k + v^j \partial_j v^k = - \vec\nabla q
\label{newf}
\end{equation}
where
\begin{equation}
q(\vec{x}, t) = - {\nabla^2 \sqrt{\rho} \over 2 m^2 \sqrt{\rho}}~~~\ .
\label{q}
\end{equation}
Following the motion, the stress tensor is
\begin{equation}
T_{jk} = \rho v_j v_k +
{1 \over 4 m^2}({1 \over \rho} \partial_j \rho \partial_k \rho
- \delta_{jk} \nabla^2 \rho)~~~~\ .
\label{stress}
\end{equation}
Comparison with Eqs.~(\ref{eom}) and (\ref{sem}) shows that
axion BEC differs from CDM: the last terms on the RHS of 
Eqs.~(\ref{newf}) and (\ref{stress}) are absent in the CDM
case.  

The extra terms are due to the Heisenberg uncertainty principle.
When one attempts to localize the axion BEC within a region of 
size $D$, the axions acquire a minimum momentum spread of order 
$\Delta p \sim \hbar/D$ and hence a velocity dispersion
$\Delta v \sim \hbar/m D$.  The axion BEC tends therefore to 
delocalize.  The tendency to delocalize is described by the 
force per unit mass $~ - \vec\nabla q$ appearing in Eq.~(\ref{newf}), 
and by the extra stresses in Eq.~(\ref{stress}).  CDM has no such 
tendency to delocalize.   Whether this difference between axion BEC 
and CDM is of observational relevance depends on the axion mass $m$.  
The properties of CDM have been observed only on very large scales, 
$D \gtrsim$ 100 kpc.  The associated velocity dispersion is tiny, 
of order $3 \cdot 10^{-24}~c~\left({10^{-5}~{\rm eV} \over m} \right)$, 
for the axion masses of interest to us.  Hence there is no distinction 
between axion BEC and CDM, on scales of observational interest, which 
follows from the extra delocalizing forces on the axion BEC.  As an 
illustration, consider the equation for the evolution of axion BEC 
density perturbations in the matter dominated phase of the expanding 
universe \cite{Ruff} \cite{CABEC,Hwang}:
\begin{equation}
\partial_t^2 \delta + 2 H \partial_t \delta
- \left(4 \pi G \rho_0 - {k^4 \over 4 m^2 a^4}\right) \delta = 0~~~\ .
\label{deneq}
\end{equation}
Here, $H \equiv \dot{a}/a$ is the Hubble expansion rate, $a(t)$ the scale 
factor, $\rho_0$ the average density of the axion BEC (assumed to be the 
dominant form of matter), $\vec{k}$ the co-moving wavevector, and $\delta$ 
the Fourier component of $(\rho - \rho_0)/\rho_0$ with wavevector $\vec{k}$.  
The last term in Eq.~(\ref{deneq}) is absent for CDM.  As a result of this 
term, whereas the Jeans length of CDM vanishes, axion BEC has Jeans length
\cite{Ruff} \cite{CABEC,Hwang}:
\begin{equation}
k_{\rm J}^{-1} = (16 \pi G \rho m^2)^{-{1 \over 4}}
= 1.02 \cdot 10^{14}~{\rm cm}
\left({10^{-5}~{\rm eV} \over m}\right)^{1 \over 2}
\left({10^{-29}~{\rm g/cm^3} \over \rho}\right)^{1 \over 4}~\ .
\label{Jeans}
\end{equation}
It is small compared to the smallest scales ($\sim$ 100 kpc)
for which we have observations on the behavior of CDM.

Several authors have proposed \cite{dmBEC} that the dark matter is a 
Bose-Einstein condensate of particles with mass of order $10^{-22}$ eV.  
When the mass is that small, the dark matter BEC behaves differently 
from CDM on scales of observational interest as a result of the 
tendency of the BEC to delocalize.  We are not considering this 
interesting possibility here because axion masses are not expected
to be so small.

In summary, cold axions with mass in the $10^{-5}$ range (give or take 
a few orders of magnitude) behave as CDM on all scales of observational 
interest in two different cases: 1) when they are decoupled and therefore
behave as a classical field, and 2) when they are all in the same particle 
state.  The latter result constrains the conditions under which axion BEC 
may differ from CDM.  For the axion BEC to differ from CDM it must be 
rethermalizing, i.e. the state that most axions are in must be changing 
in time, as when it tracks the lowest energy available state.

\section{Axion field dynamics}

We consider two types of processes through which dark matter axions may 
thermalize in the early universe: self-interactions of the $\lambda \phi^4$
type and gravitational self-interactions.  We will see in Section V, confirming
the results of ref. \cite{CABEC}, that the $\lambda \phi^4$ interaction is 
barely effective in thermalizing axions for a brief period just after they 
are produced during the QCD phase transition, whereas gravitational 
self-interactions are clearly effective in thermalizing axions after 
the photon temperature reaches approximately 
500 eV $\left({f_a \over 10^{12}~{\rm GeV}}\right)^{1 \over 2}$. 
When the axions thermalize, they form a Bose-Einstein condensate.  

In this Section, we first express the axion field dynamics in terms 
of a set of coupled oscillators.  Next we obtain the evolution equations 
for the oscillator occupation numbers up to second order in perturbation 
theory. We show that the second order terms yield the usual Boltzmann 
equation for the evolution of the momentum distribution provided that the 
transition rate between momentum states is small compared to their spread 
in energy: $\Gamma << \delta \omega$.  This latter condition defines the 
`particle kinetic regime'.  In the particle kinetic regime, the first order 
terms in the evolution equations are irrelevant because they average out in 
time.  Dark matter axions are, however, in the opposite regime: 
$\delta \omega << \Gamma$, which we refer to as the `condensed regime'.  
In the condensed regime, the first order terms do not average out in time 
and dominate over the second order terms.  Using the first order equations, 
we give expressions estimating the relaxation rate of the axion dark matter 
momentum distribution due to $\lambda \phi^4$ self-interactions and to 
gravitational self-interactions.

\subsection{Axion interactions}

Including lowest order self-interactions but neglecting gravity, the 
action density of axions is 
\begin{equation}
{\cal L}_a = - {1 \over 2} \partial_\mu \phi \partial^\mu \phi 
- {1 \over 2} m^2 \phi^2 + {\lambda \over 4!} \phi^4 + ... ~~~~\ .
\label{axact}
\end{equation}
The dots represent interactions of the axion with other particles 
and axion self-interactions which are higher order in an expansion
in powers of $\phi$.  For the axion that solves the strong CP problem 
\cite{PQ,SW,FW}, the mass $m$ and self-coupling $\lambda$ are given by 
\begin{eqnarray}
m &=& {m_\pi f_\pi \over f_a} 
{\sqrt{m_u~m_d} \over m_u + m_d} \simeq 
6 \cdot 10^{-6} {\rm eV}~{10^{12}~{\rm GeV} \over f_a}\nonumber\\
\lambda &=& {m^2 \over f_a^2} 
{m_u^3 + m_d^3 \over (m_u + m_d)^3} \simeq
0.35~{m^2 \over f_a^2}
\label{ml}
\end{eqnarray}
where $m_\pi$ is the pion mass, $f_\pi \simeq 93$ MeV the 
pion decay constant, and $m_u$ and $m_d$ are the up and 
down quark masses.  The formula for the axion mass \cite{SW} is 
obtained by expanding the effective potential for pions and axions 
to second order in the physical axion field.  To obtain $\lambda$, 
simply expand to fourth order.  

We introduce a cubic box of volume $V = L^3$ with periodic boundary 
conditions at its surface.  Inside the box, the axion field $\phi(\vec{x}, t)$ 
and its canonical conjugate field $\pi(\vec{x}, t)$ are expanded into Fourier 
components
\begin{eqnarray} 
\phi(\vec{x}, t) &=& \sum_{\vec{n}}
\left(a_{\vec n}(t)~\Phi_{\vec n}(\vec {x})~+~
a_{\vec n}^\dagger(t)~\Phi_{\vec n}^*(\vec{x})\right) \nonumber\\
\pi(\vec{x}, t) &=& \sum_{\vec{n}} (- i \omega_{\vec n})
\left(a_{\vec n}(t)~\Phi_{\vec n}(\vec {x})~-~
a_{\vec n}^\dagger(t)~\Phi_{\vec n}^*(\vec{x})\right)
\label{expansion}
\end{eqnarray}
where 
\begin{equation}
\Phi_{\vec n}(\vec{x}) = {1 \over \sqrt{2 \omega_{\vec n} V}}
~e^{i \vec{p}_{\vec n} \cdot \vec{x}}~~~~~\ ,
\label{modefct}
\end{equation}
and $\vec{n} = (n_1, n_2, n_3)$ with $n_k~(k = 1,2,3)$ 
integers, $\vec{p}_{\vec n} = {2 \pi \over L} \vec{n}$, 
and $\omega = \sqrt{\vec{p} \cdot \vec{p} + m^2}$.  The 
$a_{\vec n}$ and $a_{\vec n}^\dagger$ satisfy canonical 
equal-time commutation relations:
\begin{equation}
[a_{\vec n}(t), a_{\vec n^\prime}^\dagger (t)] = 
\delta_{{\vec n},{\vec n^\prime}}~~,~~
[a_{\vec n}(t), a_{\vec n^\prime}(t)] = 0~~\ .
\label{can}
\end{equation}
Note that we are quantizing in the Heisenberg picture, not 
the interacting picture.  

Provided the axions are non-relativistic, the Hamiltonian is
\begin{equation}
H~=~\sum_{\vec n} \omega_{\vec n}~a_{\vec n}^\dagger a_{\vec n}
~~+ \sum_{\vec{n}_1,\vec{n}_2,\vec{n}_3,\vec{n}_4}
{1 \over 4}~\Lambda_{s~\vec{n}_1,\vec{n}_2}^{~~\vec{n}_3,\vec{n}_4}~
a_{\vec{n}_1}^\dagger a_{\vec{n}_2}^\dagger a_{\vec{n}_3} a_{\vec{n}_4}
\label{axHamil}
\end{equation}
where 
\begin{equation}
\Lambda_{s~\vec{n}_1,\vec{n}_2}^{~~\vec{n}_3,\vec{n}_4}
= - {\lambda \over 4 m^2 V}~ 
\delta_{\vec{n}_1 + \vec{n}_2, \vec{n}_3 + \vec{n}_4}~~~\ .
\label{selfc}
\end{equation}
The presence of the Kronecker symbol 
$\delta_{\vec{n}_1 + \vec{n}_2, \vec{n}_3 + \vec{n}_4}$ 
expresses momentum conservation for each individual interaction. 
In Eq.~(\ref{axHamil}) we dropped all terms of the form 
$a^\dagger a^\dagger a^\dagger a^\dagger$, 
$a~a^\dagger~a^\dagger~a^\dagger$, $a~a~a~a$, and  
$a~a~a~a^\dagger$.  We are justified in doing so because 
energy conservation allows only axion number conserving 
processes at tree level.  Axion number violating processes 
occur in loop diagrams but can be safely ignored because 
they are higher order in an expansion in powers of 
${1 \over f_a}$.  In fact, all axion number violating 
processes, including the axion decay to two photons, occur 
on time scales much longer than the age of the universe in 
the axion mass range ($10^{-5}$ eV) of interest.  

In the Newtonian limit, the gravitational self-interactions of 
the axion fluid are described by 
\begin{equation}
H_g = - {G \over 2} \int d^3 x~d^3 x^\prime~ 
{\rho(\vec{x}, t) \rho(\vec{x}^\prime, t) \over 
|\vec{x} - \vec{x}^\prime|}
\label{gravint}
\end{equation}
where $\rho = {1 \over 2}(\pi^2 + m^2 \phi^2)$ is the 
axion energy density.  Because we neglect general relativistic 
corrections, our conclusions are applicable only for processes 
well within the horizon.  Substituting $\phi$ and $\pi$ by their 
expansions in terms of creation and annihilation operators, 
Eqs.~(\ref{expansion}), and dropping again all axion number 
violating terms, Eq.~(\ref{gravint}) becomes 
\begin{equation}
H_g =  \sum_{\vec{n}_1,\vec{n}_2,\vec{n}_3,\vec{n}_4}
{1 \over 4}~\Lambda_{g~\vec{n}_1,\vec{n}_2}^{~~\vec{n}_3,\vec{n}_4}~
a_{\vec{n}_1}^\dagger a_{\vec{n}_2}^\dagger a_{\vec{n}_3} a_{\vec{n}_4}
\label{gravint2}
\end{equation}
where
\begin{equation}
\Lambda_{g~\vec{n}_1,\vec{n}_2}^{~~\vec{n}_3,\vec{n}_4}
= - {4 \pi G m^2 \over V}
\delta_{\vec{n}_1 + \vec{n}_2, \vec{n}_3 + \vec{n}_4}~
\left({1 \over |\vec{p}_{\vec{n}_1} - \vec{p}_{\vec{n}_3}|^2}
+ {1 \over |\vec{p}_{\vec{n}_1} - \vec{p}_{\vec{n}_4}|^2}\right)
~~\ .
\label{gravc}
\end{equation}
Having expressed the dynamics of the axion field in terms of
coupled oscillators, we now study the time evolution of such 
systems.

\subsection{Evolution equations}

We just saw that the axion field is equivalent to a large number 
$M$ of coupled oscillators with Hamiltonian of the form
\begin{equation} 
H = \sum_{j = 1}^M~\omega_j a_j^\dagger a_j~+~
\sum_{i,j,k,l}~{1 \over 4}~\Lambda_{kl}^{ij}~
a_k^\dagger a_l^\dagger a_i a_j~~~\ .
\label{cosc}
\end{equation} 
In particular, the total number of quanta 
\begin{equation}
N = \sum_{j = 1}^M a_j^\dagger a_j
\label{totnum}
\end{equation}
is conserved.  In Eq.~(\ref{cosc}), $\Lambda_{kl}^{ij} = 
\Lambda_{lk}^{ij} = \Lambda_{kl}^{ji} = \Lambda_{ij}^{kl~*}$.
The question of interest now is the following: starting with 
an arbitrary initial state, how quickly will the averages
$\langle {\cal N}_k \rangle$ of the oscillator occupation numbers 
${\cal N}_k = a_k^\dagger a_k$ approach a thermal distribution?  
The usual approach to this question uses the Boltzmann equation.  
However, we will see that the assumptions underlying the 
Boltzmann equation are not valid for the cold axion fluid.  
So we need a more general approach.

It is instructive to start with a system of just four 
oscillators ($M$ = 4) and one interaction between them:
\begin{equation}
H = \sum_{j = 1}^4 \omega_j a_j^\dagger a_j
+ \Lambda (a_1^\dagger a_2^\dagger a_3 a_4 +
a_3^\dagger a_4^\dagger a_1 a_2)~~~\ .
\label{toyH}
\end{equation}
We have in this case
\begin{equation}
\dot{a}_1 = i [H, a_1] = i ( - \omega_1 a_1 - \Lambda a_2^\dagger a_3 a_4)
\label{adot}
\end{equation}
and therefore
\begin{equation}
\dot{\cal N}_1 = i \Lambda (a_1 a_2 a_3^\dagger a_4^\dagger - 
a_1^\dagger a_2^\dagger a_3 a_4)
\label{Ndot}
\end{equation}
and similar equations for the other $\dot{a}_j$ and $\dot{\cal N}_j$.  
We solve the equations perturbatively up to ${\cal O}(\Lambda^2)$.  
Let us define 
\begin{equation}
a_j(t) = (A_j + B_j(t)) e^{- i \omega_j t} + {\cal O}(\Lambda^2)
\label{define}
\end{equation}
where $A_j \equiv a_j(0)$ and $B_j(t)$ are respectively zeroth and 
first order, and $B_j(0) = 0$.  Eq.~(\ref{adot}) implies
\begin{equation}
\dot{B}_1 = - i \Lambda A_2^\dagger A_3 A_4 e^{+ i \Omega t} 
+ {\cal O}(\Lambda^2)~~~~\ ,
\label{Bdot}
\end{equation}
with $\Omega \equiv \omega_1 + \omega_2 - \omega_3 - \omega_4$, 
and therefore 
\begin{equation}
B_1(t) = - i \Lambda A_2^\dagger A_3 A_4~
e^{+ i \Omega t / 2}~{2 \over \Omega} \sin({\Omega t \over 2})
+ {\cal O}(\Lambda^2)~~~\ .
\label{a1t}
\end{equation}
Substituting this into Eq.~(\ref{Ndot}), we have
\begin{eqnarray}
\dot{\cal N}_1 &=& 
i \Lambda (A_1 A_2 A_3^\dagger A_4^\dagger e^{- i \Omega t} 
- h.c.)\nonumber\\
&+& \Lambda^2 [(A_2^\dagger A_2 A_3 A_3^\dagger A_4 A_4^\dagger
+ A_1 A_1^\dagger A_3 A_3^\dagger A_4 A_4^\dagger\nonumber\\
&-& A_1 A_1^\dagger A_2 A_2^\dagger A_4 A_4^\dagger
- A_1 A_1^\dagger A_2 A_2^\dagger A_3^\dagger A_3)
e^{- i \Omega t / 2}~{2 \over \Omega} \sin({\Omega t \over 2})
+ h.c.]~+~{\cal O}(\Lambda^3)\ .
\label{almost} 
\end{eqnarray}
Eq.~(\ref{almost}) may be recast in the form
\begin{eqnarray}
\dot{\cal N}_1 &=& 
i \Lambda (A_1 A_2 A_3^\dagger A_4^\dagger e^{- i \Omega t} 
- h.c.)\nonumber\\
&+& \Lambda^2 \left[{\cal N}_3 {\cal N}_4 ({\cal N}_1 + 1) ({\cal N}_2 + 1) 
- {\cal N}_1 {\cal N}_2 ({\cal N}_3 + 1) ({\cal N}_4 + 1) \right] 
{2 \over \Omega} \sin(\Omega t)
~+~{\cal O}(\Lambda^3)~~~
\label{there}
\end{eqnarray}
by rewriting the second order terms.

We now generalize to a system with an arbitrarily large number $M$ of 
coupled oscillators, Eqs.~(\ref{cosc}).  The calculation is essentially 
the same as for the $M=4$ toy model, except that there is a multiplicity 
of interaction terms to keep track of.  One finds ($l = 1 ... M$) 
\begin{eqnarray} 
\dot{\cal N}_l &=& i \sum_{i,j,k = 1}^M {1 \over 2}
(\Lambda_{ij}^{kl} A_i^\dagger A_j^\dagger A_k A_l e^{- i \Omega_{ij}^{kl} t}
 - h.c.)\nonumber\\ 
&+& \sum_{k,i,j= 1}^M {1 \over 2} |\Lambda_{ij}^{kl}|^2 
\left[{\cal N}_i {\cal N}_j ({\cal N}_l + 1)({\cal N}_k + 1) 
- {\cal N}_l {\cal N}_k ({\cal N}_i + 1)({\cal N}_j + 1)\right]
{2 \over \Omega_{ij}^{kl}} \sin(\Omega_{ij}^{kl} t)
\nonumber\\ 
&+& \sum_{k,i,j= 1}^M \sum_{{p,m,n=1} \atop (p;m,n)\neq(k;i,j)}^M 
[{1 \over 2} \Lambda_{kl}^{ij}
\Lambda_{mn}^{lp} A_m^\dagger A_n^\dagger A_k^\dagger A_p A_i A_j
e^{i (\Omega_{ij}^{kl} + \Omega_{lp}^{mn}/2)t}~
{1 \over \Omega_{lp}^{mn}} \sin({\Omega_{lp}^{mn} \over 2} t) 
+ h.c. ]\nonumber\\
&+& \sum_{k,i,j= 1}^M \sum_{{p,m,n=1} \atop (p;m,n)\neq(l;i,j)}^M 
[{1 \over 2} \Lambda_{kl}^{ij}
\Lambda_{mn}^{kp} A_l^\dagger A_m^\dagger A_n^\dagger A_p A_i A_j 
e^{i (\Omega_{ij}^{kl} + \Omega_{kp}^{mn}/2)t}~
{1 \over \Omega_{kp}^{mn}} \sin({\Omega_{kp}^{mn} \over 2} t) 
+ h.c. ]\nonumber\\
&-& \sum_{k,i,j= 1}^M \sum_{{p,m,n=1} \atop (p;m,n) \neq (j;l,k)}^M 
[{1 \over 2} \Lambda_{lk}^{ij}
\Lambda_{ip}^{mn} A_l^\dagger A_k^\dagger A_p^\dagger A_m A_n A_j
e^{i (\Omega_{ij}^{kl} + \Omega_{mn}^{ip}/2)t}~
{1 \over \Omega_{mn}^{ip}} \sin({\Omega_{mn}^{ip} \over 2} t) +
h.c. ]\nonumber\\
&-&  \sum_{k,i,j = 1}^M \sum_{{p,m,n =1} \atop (p;m,n) \neq (i;l,k)}^M 
[{1 \over 2} \Lambda_{lk}^{ij}
\Lambda_{jp}^{mn} A_l^\dagger A_k^\dagger A_p^\dagger A_i A_m A_n 
e^{i (\Omega_{ij}^{kl} + \Omega_{mn}^{jp}/2)t}~
{1 \over \Omega_{mn}^{jp}} \sin({\Omega_{mn}^{jp} \over 2} t) 
+ h.c. ]\nonumber\\
&+& {\cal O}(\Lambda^3)~~~\ ,
\label{full}
\end{eqnarray}
where $\Omega_{ij}^{kl} \equiv \omega_k + \omega_l - \omega_i - \omega_j$.
The double sums are absent in the toy model because there is only one 
interaction in that case.  At any rate, the double sums will not play 
an important role in the discussion that follows.

\subsection{The particle kinetic regime}

The interaction terms cause transitions in which a quantum is moved 
from oscillator $j$ to oscillator $k$ and at the same time a quantum 
is moved from oscillator $i$ to oscillator $l$.  As a general rule, 
energy is not conserved in any one transition.  Only the total energy 
of the fluid is conserved.  However, for the oscillator couplings of the 
axion field case, Eqs.~(\ref{selfc}) and (\ref{gravc}), three-momentum 
is conserved in each transition.

In almost all systems of physical interest, the rate at which the 
occupation number of a typical system oscillator changes is small 
compared to the energy exchanged in the transitions it makes, i.e. 
$\Omega_{ij}^{kl} t >> 1$. This condition defines the `particle 
kinetic regime'.  In the particle kinetic regime, the first order 
terms in Eq.~(\ref{full}), as well as the second order terms in 
the double sums, average to zero in time.  In addition, energy 
is conserved in each transition because
\begin{equation}
{2 \over \Omega_{ij}^{kl}} \sin(\Omega_{ij}^{kl}~t)
\rightarrow 2 \pi \delta(\Omega_{ij}^{kl})
\label{encon}
\end{equation}
for $\Omega_{ij}^{kl}~t \rightarrow \infty$.  We have then 
\begin{equation}
<\dot{\cal N}_l> =   
+ \sum_{i,j,k = 1}^M {1 \over 2} |\Lambda_{ij}^{kl}|^2
\left[{\cal N}_i {\cal N}_j ({\cal N}_l + 1)({\cal N}_k + 1) 
- {\cal N}_l {\cal N}_k ({\cal N}_i + 1)({\cal N}_j + 1)\right]
2 \pi \delta(\Omega_{ij}^{kl})~~~\ .
\label{ludwig}
\end{equation}
Note that the average on the LHS of this equation is a time 
average, not a quantum-mechanical average.  Eq.~(\ref{ludwig})
is valid as an operator statement.

If we substitute for the oscillator couplings those, Eq.~(\ref{selfc}), 
implied by $\lambda \phi^4$ self-interactions and take the infinite 
volume limit, we recover the Boltzmann equation \cite{Semi} as an 
operator statement:
\begin{eqnarray}
<\dot{\cal N}_1> &=& {1 \over 2 \omega_1} 
\int {d^3 p_2 \over (2 \pi)^3 2 \omega_2}
\int {d^3 p_3 \over (2 \pi)^3 2 \omega_3}
\int {d^3 p_4 \over (2 \pi)^3 2 \omega_4}
\nonumber\\ 
&\lambda^2&(2 \pi)^4 
\delta^4(p_1 + p_2 - p_3 - p_4)~{1 \over 2}~
\nonumber\\
&~&[({\cal N}_1 + 1) ({\cal N}_2 + 1) {\cal N}_3 {\cal N}_4 
- {\cal N}_1 {\cal N}_2 ({\cal N}_3 + 1) ({\cal N}_4 + 1)]
\label{Boltz} 
\end{eqnarray}
including the factor (explicitly shown) of 1/2 for identical particles 
in the final state.  In Eq.~(\ref{Boltz}), ${\cal N}_k$ is short for 
${\cal N}_{{\vec p}_k}$.   The $a + a \rightarrow a + a$ scattering 
cross-section due to the $\lambda \phi^4$ self-interaction is  
\begin{eqnarray}
\sigma_\lambda &=& {1 \over |\vec{v}_1 - \vec{v}_2|} {1 \over 2 \omega_1}
{1 \over 2 \omega_2} \int {d^3 p_3 \over (2 \pi)^3 2 \omega_3}
 \int {d^3 p_4 \over (2 \pi)^3 2 \omega_4} \lambda^2 {1 \over 2}
(2 \pi)^4 \delta^4(p_1 + p_2 - p_3 - p_4)\nonumber\\
&=& {\lambda^2 \over 64 \pi} {1 \over m^2}
\label{xsec}
\end{eqnarray}
where the last equality holds in the non-relativistic limit.  The particle 
density in physical space is
\begin{equation}
n = \int {d^3 p \over (2 \pi)^3} {\cal N}_{\vec p}~~~\ .
\label{parden}
\end{equation}
If most states are not occupied, Eqs.~(\ref{Boltz}),(\ref{xsec}) 
and (\ref{parden}) imply the usual expression for the relaxation rate
\begin{equation}
\Gamma \sim {\dot{\cal N} \over {\cal N}} \sim n~\sigma~\delta v 
\label{relax1}
\end{equation}
where $\delta v$ is a measure of the velocity dispersion in the 
fluid.  

However, in the case of cold dark matter axions, the momentum states 
that are occupied are enormously occupied \cite{CABEC}: 
${\cal N}_{\vec p} \sim 10^{61}
~\left({f \over 10^{12}~{\rm GeV}}\right)^{8 \over 3}$
for $p \lesssim p_{\rm max}(t) \sim {1 \over t_1}{a(t_1) \over a(t)}$; 
see Section V.   For such a situation, consider the factor 
\begin{equation}
F({\cal N}) \equiv ({\cal N}_1 + 1) ({\cal N}_2 + 1) {\cal N}_3 {\cal N}_4
- {\cal N}_1 {\cal N}_2 ({\cal N}_3 + 1) ({\cal N}_4 + 1) 
\label{FN}
\end{equation}
that appears in Eq.~(\ref{Boltz}).  The first term in $F({\cal N})$ 
multiplies the contribution to $<\dot{\cal N}_1>$ from 
$a(p_1) + a(p_2) \leftarrow a(p_3) + a(p_4)$ transitions 
whereas the second terms multiplies the contribution from  
$a(p_1) + a(p_2) \rightarrow a(p_3) + a(p_4)$.   By energy 
conservation, if the initial particle states (e.g. $\vec{p}_1$ 
and $\vec{p}_2$) are both highly occupied, the final particle 
states ($\vec{p}_3$ and $\vec{p}_4$) are almost always highly 
occupied as well.  The RHS of Eq.~(\ref{Boltz}) is dominated 
by regions of momentum space where all four ${\cal N}_k$ are 
of order ${\cal N} \sim 10^{61}$.  The rate for 
$a(p_1) + a(p_2) \rightarrow a(p_3) + a(p_4)$ 
transitions, per pair of initial state particles, is multiplied by 
$({\cal N}_3 + 1) ({\cal N}_4 + 1) \sim {\cal N}^2$ compared to the 
case where most states are empty.  However, the rate for 
$a(p_1) + a(p_2) \leftarrow a(p_3) + a(p_4)$ is similarly multiplied 
by order ${\cal N}^2$.  The two cancel each other since the terms of 
order ${\cal N}^4$ cancel out in $F({\cal N}$).  However, the order 
${\cal N}^3$ terms do not cancel out.  The upshot is that, although 
the rates at which quanta enter and leave any one state ${\vec p}_1$ 
are both enhanced factors of ${\cal N}^2$, the difference between 
those two rates, i.e. the relaxation rate, is multiplied by one 
factor of ${\cal N}$:
\begin{equation}
\Gamma \sim {\dot{\cal N} \over {\cal N}} 
\sim n~\sigma~\delta v~{\cal N}~~~\ .
\label{relax2}
\end{equation}
Using this equation, we show in Section V that cold axions  
barely thermalize near time $t_1$, just after they are 
produced during the QCD phase transition.  However the 
condition $\Gamma << \delta \omega$ that defines the 
particle kinetic regime is only borderline satisfied at 
time $t_1$ and is violated afterward.  Shortly after $t_1$ 
the axion fluid enters the condensed regime.  Relaxation in 
the condensed regime is discussed in the next subsection.

We would like to conclude this subsection with remarks about
gravitational interactions.  It has long been recognized that 
gravitational interactions do not fit into the usual discussion 
of thermalization \cite{Kad}. Indeed the total cross section for 
gravitational scattering is formally infinite, and Eqs.~(\ref{relax1}) 
and (\ref{relax2}) are clearly invalid.  Let's use our derivation of the 
Boltzmann equation to shed some light on this matter.  Eq.~(\ref{full}) is 
valid for all interactions since it only assumes the ordinary rules of 
quantum mechanics.  However, Eq.~(\ref{ludwig}) is valid only in the 
particle kinetic regime.  It does not apply to transitions with very 
small momentum, and hence very small energy, transfer since the particle 
kinetic condition $\Omega_{ij}^{kl}~t >> 1$ is violated then.  The 
divergence in the cross-section for gravitational scattering is at 
small momentum transfers (forward scattering), where the assumptions 
underlying the calculation of the cross-section are invalid.  Even 
the notion of scattering is dubious there.  We should restrict the 
sum in Eq.~(\ref{ludwig}) to transitions that are in the particle 
kinetic regime.  This will cut off the divergence in the total 
cross section for gravitational scattering but the differential 
cross section will still be peaked in the forward direction.  
Forward scattering has only limited effectiveness in relaxing 
a momentum distribution since the particle momenta change only 
a little bit at the time.  The cross section appearing in  
Eqs.~(\ref{relax1}) and (\ref{relax2}) is the cross-section for 
large angle scattering since only large angle scattering has an 
immediate effect on the momentum distribution.   The issue does 
not arise for $\lambda \phi^4$ interactions because the differential 
cross-section is not sharply peaked in the forward direction in that 
case.  We conclude that one may use Eq.~(\ref{relax1}) or (\ref{relax2}) 
with 
\begin{equation}
\sigma_g \sim {4 G^2 m^2 \over (\delta v)^4}
\label{lags}
\end{equation}
to estimate the relaxation rate by gravitational scattering in the 
particle kinetic regime.

\subsection{The condensed regime}

Consider a huge number $N$ of particles occupying a small number $K$ of 
states.  The average occupation number of those states that are highly 
occupied is ${\cal N} = {N \over K} >> 1$.  Assume that the energy spread 
of the highly occupied states is small compared to the evolution rate (to 
be determined) of the system:  $\delta \omega << \Gamma$.  For transitions 
between such closely spaced states, 
\begin{equation}
e^{- i \Omega_{ij}^{kl} t} = 1
\label{one}
\end{equation}
in Eq.~(\ref{full}).  The first order terms in that equation no longer 
average to zero.  We want to estimate the relaxation rate under these 
conditions.

It may appear at first that we are embarking on a meaningless quest because 
the condition  $\Omega_{ij}^{kl} t << 1$ implies that the energy difference
between highly occupied states is too small for anyone to tell whether a 
particle has made a transition from one highly occupied state to another 
in time $t$.  That is indeed so.  However, for the case of interest to 
us, $(\Omega_{ij}^{kl} t) {\cal N} >> 1$ and it is therefore meaningful to 
ask whether of order ${\cal N}$ particles have made a transition between 
highly occupied states. The latter is the question which is relevant to 
estimating the relaxation rate. 

Let us define $c_l(t) \equiv a_l(t) e^{i \omega_l t}$.  The Hamiltonian
(\ref{cosc}) implies
\begin{equation} 
\dot{c}_l(t) = - i \sum_{k,i,j = 1}^M {1 \over 2}~
\Lambda_{kl}^{ij}~c_k^\dagger c_i c_j e^{i \Omega_{ij}^{kl} t}~~~~\ .
\label{foeq}
\end{equation}
Let us define further 
\begin{equation}
c_l(t) \equiv C_l(t) + d_l(t)
\label{Clt}
\end{equation}
where the $d_l(t)$, like the $c_l(t)$, are annihilation operators satisfying
canonical equal time commutation relations and the $C_l(t)$ are complex 
c-number functions which satisfy the classical equations of motion 
\begin{equation}
\dot{C}_l(t) = - i \sum_{k,i,j = 1}^M {1 \over 2}
\Lambda_{kl}^{ij}~C_k^* C_i C_j e^{i \Omega_{ij}^{kl} t}~~~~\ .  
\label{foeq2}
\end{equation}
For the highly occupied cold axion states, the $C_l$ have magnitude of 
order $\sqrt{\cal N}$.  The relaxation rate of the highly condensed 
cold axions is the inverse of the time scale over which those $C_l(t)$ 
change by order $\sqrt{\cal N}$.  

Since the sum in Eq.~(\ref{foeq2}) is dominated by terms for which 
$k$, $i$ and $j$ label highly occupied axion states,
\begin{equation}
\dot{C}_l(t) \sim - i \sum_{k,i,j = 1}^K {1 \over 2}
\Lambda_{kl}^{ij}~C_k^* C_i C_j~~~~\ .
\label{foeq3}  
\end{equation}
For $\lambda \phi^4$ self-interactions, we substitute Eq.~(\ref{selfc}).  
This yields
\begin{equation}
\dot{C}_{\vec {p}_1}(t) \sim + i {\lambda \over 4 m^2 V}
\sum_{\vec{p}_2, \vec{p}_3} {1 \over 2}
C_{\vec{p}_2}^* C_{\vec{p}_3} C_{\vec{p}_4} 
\label{foeq4} 
\end{equation}
where $\vec{p}_4 = \vec{p}_1 + \vec{p}_2 - \vec{p}_3$, and the 
sum is restricted to the $K$ highly occupied states for which 
$p \lesssim p_{\rm max}$.  We may think of the terms on the RHS 
of Eq.~(\ref{foeq4}) as steps in a random walk in complex space.  
The magnitude of each step is of order ${\cal N}^{3 \over 2}$ and 
the number of steps is of order $K^2$.  Hence 
\begin{equation}
\dot{C}_{\vec p} \sim {\lambda \over 4 m^2 V} K {\cal N}^{3 \over 2}
\sim {\lambda \over 4 m^2 V} N {\cal N}^{1 \over 2}~~~\ .
\label{foeq5}
\end{equation}
Hence our estimate for the relaxation rate due to $\lambda \phi^4$ 
self-interactions in the condensed regime \cite{CABEC}:
\begin{equation}
\Gamma_\lambda \sim {1 \over 4} n \lambda m^{-2}
\label{relcon}
\end{equation}
where $n = N/V$ is the density of particles in the highly 
occupied closely spaced states.  Likewise, using Eq.~(\ref{gravc}), 
we find that through gravitational self-interactions 
\begin{equation}
\dot{C}_{\vec {p}_1}(t) \sim + i {4 \pi G m^2 \over V}
\sum_{\vec{p}_2, \vec{p}_3} {1 \over 2}
C_{\vec{p}_2}^* C_{\vec{p}_3} C_{\vec{p}_4} 
\left({1 \over |\vec{p}_1 - \vec{p}_3|^2}
+ {1 \over |\vec{p}_1 - \vec{p}_4|^2}\right)~~~\ .
\label{foeq4g}
\end{equation}  
The corresponding relaxation rate is
\begin{equation}
\Gamma_g \sim 4 \pi G n m^2 \ell^2
\label{relcong}
\end{equation}
where $\ell \sim {1 \over p_{\rm max}}$ is the correlation length 
of the particles.  

We note that at the boundary between the particle kinetic and 
condensed regimes, where $\delta \omega \sim \Gamma$, the two 
estimates of the relaxation rate agree with one another.  
Indeed, at that boundary,
\begin{equation}
\delta v ~{\cal N} \sim \delta v {n \over (\delta p)^3} \sim 
{n \over m^2 \delta \omega} \sim {n \over m^2 \Gamma}~~~\ .
\label{border}
\end{equation}
Substituting this into Eq.~(\ref{relax2}) for the relaxation 
rate due to $\lambda \phi^4$ self-interactions in the particle 
kinetic regime yields Eq.~(\ref{relcon}) which is the corresponding
estimate in the condensed regime.  The same holds true for the 
gravitational self-interactions.

Let us also note that Eqs.~(\ref{relcon}) and (\ref{relcong}) are not 
valid when almost all axions are in a single state ($K = 1$), as when the 
Bose-Einstein condensation has been completed.  Indeed, if $K=1$, there is 
only one term in the sum on the RHS of Eqs.~(\ref{foeq4}) and (\ref{foeq4g}) 
that is enhanced by large occupation numbers, i.e. the term for which both 
$\vec{p}_2$ and $\vec{p}_3$ equal the momentum of the single highly occupied 
state, and it describes an interaction with zero momentum transfer.  Thus, 
once the Bose-Einstein condensation is complete and all axions are in the 
lowest energy state, any further thermalization is suppressed.

Finally consider transitions $a(\vec{p}_1) + a(\vec{p}_2) 
\leftrightarrow a(\vec{p}_3) + a(\vec{p}_4)$ where $\vec{p}_2$ and 
$\vec{p}_4$ are momenta of highly occupied states but $\vec{p}_1$
and $\vec{p}_3$ are not.  Such transitions are in the condensed 
regime because the momentum transfer, and hence the energy transfer, 
is small.  Eqs.~(\ref{foeq4}) and (\ref{foeq4g}) apply to such transitions 
and imply that the rate at which states with $p > p_{\rm max}$ modify their 
occupation numbers is also given by Eqs.~(\ref{relcon}) and (\ref{relcong}) 
with the proviso that the quanta can only move between states differing in 
momentum by less than $p_{\rm max}$.  

Eq.~(\ref{relcong}) has a simple interpretation. The axions, having 
energy density $m n$ and correlation length $\ell$ produce gravitational 
fields $g \sim 4 \pi G \rho \ell$.  The rate at which these gravitational 
fields modify the momentum of a particle by an amount $\Delta p$ is of order
\begin{equation}
\Gamma_g \sim {g \omega \over \Delta p} 
\sim 4 \pi G n m \ell {\omega \over \Delta p}
\label{general}
\end{equation}
where $\omega$ is the energy of the particle.  For the axions themselves, 
we obtain the relaxation rate Eq.~(\ref{relcong}) by substituting  
$\omega = m$ and $\Delta p \sim \ell^{-1}$.  Eq.~(\ref{general}) shows 
that the momentum distribution of any particle species is modified by 
the gravitational fields of the cold axion fluid and therefore that 
gravitational interactions may produce thermal contact between the 
cold axions and other particle species.  This is discussed in the 
next subsection.

\subsection{Other species}

Our purpose in this subsection is to estimate the gravitational 
interaction rates of other species - baryons, relativistic axions 
and photons - with the cold axion fluid.  We are motivated by the 
question, discussed in Section V, whether these other species come 
into thermal contact with the cold axion fluid.  

The Hamiltonian describing gravitational interactions between
the cold axions and any other species has the general form:
\begin{equation} 
H = \sum_{j = 1}^M~\omega_j a_j^\dagger a_j~+~
\sum_{r = 1}^S \omega_r b_r^\dagger b_r~+~
\sum_{i,j,k,l}~{1 \over 4}~\Lambda_{kl}^{ij}~
a_k^\dagger a_l^\dagger a_i a_j~+~
\sum_{j,k,r,s}~\Lambda_{b~ks}^{jr}~a_k^\dagger b_s^\dagger a_j b_r~~~~\ ,
\label{genH}
\end{equation}
where $\Lambda_{b~ks}^{jr}= \left(\Lambda_{b~jr}^{ks}\right)^*$.
The $b_r$ are the annihilation operators for quanta of the new
species.  They satisfy canonical commutation or anti-commutation 
relations.  The $\omega_r$ are the energies of those quanta.  The 
other symbols ($\omega_j$, $a_j$ and $\Lambda_{kl}^{ij}$) have the 
same meaning as in Eq.~(\ref{cosc}).  As before, we quantize in a 
box of volume $V = L^3$ with periodic boundary conditions.  The labels 
of the new particle states are then $r = (\vec{n}, \sigma)$,  giving 
their momenta $\vec{p} = {2 \pi \over L} \vec{n}$ and their spin $\sigma$.  
Their energy is $\omega = \sqrt{\vec{p}\cdot\vec{p} + m_b^2}$ where $m_b$ 
is the mass of the new species.

We define $c_j(t) \equiv a_j(t) e^{i \omega_j t}$ as before, and 
$c_r^\prime(t) \equiv b_r(t) e^{i \omega_r t}$.  The Heisenberg
equations of motion for the $c_r^\prime(t)$ are then:
\begin{equation}
\dot{c}_s^\prime = - i \sum_{j,k,r} \Lambda_{b~ks}^{jr} 
c_k^\dagger c_j c_r^\prime e^{i \Omega_{jr}^{ks}~t}
\label{Heisp}
\end{equation}
where $\Omega_{jr}^{ks} \equiv \omega_k + \omega_s - \omega_j - \omega_r$.  
Because 3-momentum is conserved in each interaction, the $\Lambda_{b~ks}^{jr}$ 
have the form:
\begin{equation}
\Lambda_{b~ks}^{jr} = 
- \lambda_{b~ks}^{jr}~\delta_{\vec{p}_k + \vec{p}_s, \vec{p}_j + \vec{p}_r}~~\ .
\label{3mc}
\end{equation}
The important contributions in the sum on the RHS of Eq.~(\ref{Heisp}) are 
from terms in which both $j$ and $k$ label highly occupied cold axion states.
Therefore 
\begin{equation}
\dot{c}_s^\prime \sim + i \sum_{k,j = 1}^K 
\lambda_{b~ks}^{jr} C_k^* C_j c_r^\prime~e^{i(\omega_s - \omega_r) t}
\label{eqeqp}
\end{equation}
with $\vec{p}_r = \vec{p}_s + \vec{p}_k - \vec{p}_j$.  As before, $K$ 
is the number of highly occupied cold axion states and the $C_k$ are 
defined by Eq.~(\ref{Clt}).  Again, the sum on the RHS of Eq.~(\ref{eqeqp}) 
represents a random walk in complex space.  The number of steps is $K^2$ and 
the typical step size is $\lambda_b~{\cal N} c^\prime$, where $\lambda_b$ is 
the typical value of $\lambda_{b~ks}^{jr}$.  The rate at which all quanta of 
the new species may move to neighboring states seperated in momentum space by 
less than $\delta p \sim {1 \over \ell}$ is therefore
\begin{equation}
\Gamma_{b, \delta p} \sim K {\cal N} \lambda_b = \lambda_b~N
\label{litt}
\end{equation}
where $N = K {\cal N}$ is the number of cold axions in volume $V$. The
relaxation rate of the new species is then
\begin{equation}
\Gamma_b \sim \lambda_b~N~{\delta p \over \Delta p} \sim 
\lambda_b~N~{1 \over \ell \Delta p}
\label{tota}
\end{equation}
where $\Delta p$ is the momentum dispersion of the new species 
population.  (If the momentum dispersion is very different in 
the initial state than in the final state, $\Delta p$ is the 
larger of the two.)  Eq.~(\ref{tota}) assumes that the $b$ particles
are bosons or non-degenerate fermions.  If they are degenerate fermions, 
their relaxation rate is suppressed, relative to Eq.~(\ref{tota}), by 
Pauli blocking.

Also, let us reiterate that when most cold axions are in the lowest 
energy state, implying $K = 1$, thermalization is suppressed compared 
to the estimate in Eq.~(\ref{tota}), because there is only one term in 
the sum of Eq.~(\ref{eqeqp}) in that case and the momentum transfer 
vanishes for that term.

\subsubsection{Baryons}

For non-relativistic species, such as baryons and WIMPs, the 
term in the Hamiltonian that describes gravitational interactions
with the cold axions is 
\begin{equation}
H_B = - G \int d^3x~d^3x^\prime 
{\rho(\vec{x}, t) \rho_B(\vec{x}^\prime, t) \over 
|\vec{x} - \vec{x}^\prime|}~~~\ ,
\label{intb}
\end{equation} 
where 
\begin{equation}
\rho_B(\vec{x}, t) = {m_B \over V} \sum_{\vec{n}, \vec{n}^\prime, \sigma}
b_{\vec{n},\sigma}^\dagger~b_{\vec{n}^\prime,\sigma}~
e^{i(\vec{p}^{~\prime} - \vec{p})\cdot\vec{x}}~~~\ ,
\label{rhob}
\end{equation}
and $m_B$ is the mass of the non-relativistic particle.  This yields
\begin{equation}
\lambda_{B~\vec{n}_3,(\vec{n}_4,\sigma^\prime)}^{\vec{n}_1,(\vec{n}_2,\sigma)}
= + {4 \pi G m m_B \over V q^2}~
\delta_{\sigma^\prime}^\sigma~~~\ ,
\label{lamb}
\end{equation}
where $\vec{q} = \vec{p}_1 - \vec{p}_3$ is the momentum transfer.  Since 
$q \sim \ell^{-1}$, the $B$ particles have relaxation rate 
\begin{equation}
\Gamma_B \sim 4 \pi G n m m_B {\ell \over \Delta p_B}~~\ ,
\label{relb}
\end{equation}
where $\Delta p_B$ is their momentum dispersion.  Eq.~(\ref{relb}) assumes 
that the $B$ particles are bosons or non-degenerate fermions, as is the case 
for baryons and WIMPs.

\subsubsection{Hot axions}

For relativistic species, the term that describes gravitational interactions 
with the cold axion fluid is 
\begin{equation}
H_r = - \int d^3x~{1 \over 2} h_{\alpha\beta}T^{\alpha\beta}
\label{intr}
\end{equation}
where $T^{\alpha\beta}(\vec{x}, t)$ is the stress-energy-momentum 
tensor of this species and $h_{\alpha\beta}$ is the perturbation 
of the space-time metric caused by the cold axions:
\begin{eqnarray}
h_{00}(\vec{x}, t) &=& 2 G \int d^3x^\prime 
{\rho(\vec{x}^\prime, t) \over |\vec{x} - \vec{x}^\prime|}\nonumber\\
h_{0k}(\vec{x}, t) &=& 0\nonumber\\
h_{kl}(\vec{x}, t) &=& 2 G \int d^3x^\prime 
{\rho(\vec{x}, t) \over |\vec{x} - \vec{x}^\prime|^3}
(x_k - x_k^\prime)(x_l - x_l^\prime)~~~\ .
\label{metb}
\end{eqnarray}
Note that $h_\alpha^\alpha = 0$.  For a scalar field $\phi(x)$
\begin{equation}
H_r = - \int d^3x~{1 \over 2} h_{\alpha\beta}~\partial^\alpha \phi~
\partial^\beta \phi~~~\ .
\label{intrs}
\end{equation}
After some algebra, Eq.~(\ref{intrs}) yields
\begin{equation}
\lambda_{r~\vec{n}_3,\vec{n}_4}^{\vec{n}_1,\vec{n}_2}
= + {4 \pi G m \over V q^2 \sqrt{\omega_2\omega_4}}~
[\omega_2 \omega_4 + \vec{p}_2 \cdot \vec{p}_4 - 
2{(\vec{q} \cdot \vec{p}_2)(\vec{q} \cdot \vec{p}_4) \over q^2}]~~~\ ,
\label{lamr}
\end{equation}
where $\vec{q} = \vec{p}_1 - \vec{p}_3$.  The relaxation rate for 
relativistic scalars through gravitational interactions with the 
highly occupied low momentum axion modes is thus of order
\begin{equation}
\Gamma_r \sim 4 \pi G n m \ell~~~~\ ,
\label{relr}
\end{equation}
since $q \sim \ell^{-1}$ and $\Delta p \sim \omega$.

\subsubsection{Photons}

The term that describes the gravitational interactions of photons 
with the cold axion fluid is
\begin{equation}
H_\gamma = - \int d^3x {1 \over 2} h_{\alpha\beta} F^{\alpha\mu} F^\beta_{~\mu}
\label{intg}
\end{equation}
where $F_{\alpha\beta}$ is the electromagnetic field strength tensor, 
and the $h_{\alpha\beta}$ are given by Eqs.~(\ref{metb}) as before.  
This yields
\begin{equation}
\lambda_{\gamma~\vec{n}_3,(\vec{n}_4,\vec{\epsilon}_4)}
^{\vec{n}_1,(\vec{n}_2,\vec{\epsilon}_2)}
= + {8 \pi G m \over V q^4 \sqrt{\omega_2\omega_4}}~
[\omega_2 \omega_4 (\vec{\epsilon}_2\cdot\vec{q})
(\vec{\epsilon}_4^{~*} \cdot \vec{q}) +
(\vec{p}_2\times\vec{\epsilon}_2)\cdot\vec{q}~
(\vec{p}_4\times\vec{\epsilon}_4^{~*})\cdot\vec{q}]~~~\ ,
\label{lamg}
\end{equation}
where $\vec{\epsilon}_2$ and $\vec{\epsilon}_4$ are the polarization 
vectors of the initial and final state photons.  We find therefore
\begin{equation}
\Gamma_\gamma \sim  4 \pi G n m \ell
\label{relg}
\end{equation}
for the relaxation rate of photons.  It is the same as for relativistic
axions, Eq.~(\ref{relr}), in order of magnitude.

\section{Numerical simulations}

As far as we are aware, there has been no detailed discussion of thermalization 
in the condensed regime prior to this work, presumably because most many body 
systems thermalize in the opposite particle kinetic regime.  The cold axion dark
matter fluid may be the only physical system for which the condensed regime is 
the relevant one.  Since direct observation of cold axions is impossible, we 
cannot verify the validity of our estimates by empirical methods.  However, we 
may use numerical simulation as a check.  In this section we construct a toy model
which thermalizes in the condensed regime but is sufficiently simple that it can 
be simulated numerically.  The simulation agrees with our estimate of its 
thermalization rate.

The Hamiltonian for the systems of interest to us has the general form given 
in Eq.~(\ref{cosc}) which we repeat here for convenience:
\begin{equation}
H = \sum_{j = 1}^M~\omega_j a_j^\dagger a_j~+~
\sum_{i,j,k,l}~{1 \over 4}~\Lambda_{kl}^{ij}~
a_k^\dagger a_l^\dagger a_i a_j~~~\ .
\label{coscr}
\end{equation}
As before, let us define $c_j(t) \equiv a_j(t)e^{i \omega_j t}$.  The 
$c_j(t)$ satisfy the Heisenberg equations of motion
\begin{equation}
\dot{c}_l(t) = - i \sum_{k,i,j = 1}^M {1 \over 2}~
\Lambda_{kl}^{ij}~c_k^\dagger c_i c_j e^{i \Omega_{ij}^{kl} t}~~~~\ ,
\label{foeqr}
\end{equation}
where $\Omega_{ij}^{kl} \equiv \omega_k + \omega_l - \omega_i - \omega_j$.
The system is in the condensed regime when a large number of quanta are in 
states whose spread $\delta \omega$ in energy is small compared to the 
evolution rate of the system, so that $e^{i \Omega_{ij}^{kl} t} = 1$
for the dominant terms on the RHS of Eq.~(\ref{foeqr}).  We discussed
thermalization in the condensed regime in Section III.D.  Our estimate 
for the thermalization rate is 
\begin{equation}
\Gamma \sim \sqrt{I}~{\cal N}~\Lambda
\label{est}
\end{equation}
where $I$ is the number of interaction terms on the RHS of Eq.~(\ref{foeqr})
between highly occupied states, ${\cal N}$ the average occupation number of 
those states, and $\Lambda$ an average value of the $\Lambda_{kl}^{ij}$.

We may expand any state of the system 
\begin{equation}
|\Psi(t)> = \sum_{\{\cal N\}}~\Psi(\{{\cal N}\}, t)~
e^{- i E(\{{\cal N}\}) t}~|\{\cal N\}>
\label{expan}
\end{equation}
in terms of an orthonormal set of Fock space states
\begin{equation}
|\{{\cal N}\}> = |{\cal N}_1, {\cal N}_2, ... , {\cal N}_M> 
= \prod_{j=1}^M {1 \over \sqrt{{\cal N}_j !}} (a_j^\dagger)^{{\cal N}_j}~|0>~~\ .
\label{Fock}
\end{equation}
These are the eigenstates of the non-interacting Hamiltonian.  The corresponding
eigenvalues are
\begin{equation}
E(\{{\cal N}\}) = \sum_{j=1}^M {\cal N}_j \omega_j~~~~\ .
\label{eeig}
\end{equation}
The probability for the system to be in state $|\{{\cal N}\}>$ at time $t$ is 
$|\Psi(\{{\cal N}\}, t)|^2$.  

The time evolution equation 
\begin{equation}
i {\partial \over \partial t} |\Psi(t)> = H |\Psi(t)>
\label{Erwin}
\end{equation}
implies
\begin{equation}
i \dot{\Psi}(\{{\cal N}\}, t) = \sum_{i,j,k,l}~{1 \over 4} \Lambda_{kl}^{ij}
e^{- i \Omega_{kl}^{ij} t} 
\sqrt{({\cal N}_i + 1)({\cal N}_j + 1) {\cal N}_k {\cal N}_l}
~\Psi(\{_{kl}^{ij}{\cal N}\}, t) 
\label{psiev}
\end{equation}
where
\begin{eqnarray}
_{kl}^{ij}{\cal N}_p~~&=&~~{\cal N}_p~~~~~~~~~{\rm if}~p \neq i,j,k,l\nonumber\\
&=&~~{\cal N}_p + 1~~~~{\rm if}~p = i~{\rm or}~j\nonumber\\
&=&~~{\cal N}_p - 1~~~~{\rm if}~p = k~{\rm or}~l~~~\ .
\label{fancy}
\end{eqnarray}
The $\{_{kl}^{ij}{\cal N}\}$ configuration is the same as the 
$\{{\cal N}\}$ configuration except that two particles are moved, 
from the states $k$ and $l$ to the states $i$ and $j$.  In obtaining
Eqs.~(\ref{psiev}) and (\ref{fancy}) we assumed that the couplings 
$\Lambda_{kl}^{ij}$ vanish when $k=l$ or $i=j$.  If we do not impose
this restriction, Eqs.~(\ref{psiev}) and (\ref{fancy}) are a little 
more complicated.  However the complications do not change the estimate
of the thermalization rate, so we ignore them.

The evolution of the system can be determined by solving Eqs.~(\ref{psiev}) 
numerically.  For a system in the condensed regime, the important terms on 
the RHS of that equation all have $e^{- i \Omega_{kl}^{ij} t} = 1$, 
so that 
\begin{equation}
i \dot{\Psi}(\{{\cal N}\}, t) = \sum_{i,j,k,l}~{1 \over 4} \Lambda_{kl}^{ij}
\sqrt{({\cal N}_i + 1)({\cal N}_j + 1) {\cal N}_k {\cal N}_l}
~\Psi(\{_{kl}^{ij}{\cal N}\}, t)~~~\ .
\label{crev} 
\end{equation}
Recall that the cold axion fluid is in the condensed regime because 
the axions occupy relatively few states, each state is hugely occupied 
(${\cal N} \sim 10^{61}$), the particle states have a small spread in 
energy ($\delta\omega \Gamma^{-1} << 1$) yet, because 
$~\delta\omega~{\cal N} \Gamma^{-1} >> 1$, it is meaningful to ask what 
is the distribution of axions over the closely spaced states.  The multiple 
requirements are satisfied only because ${\cal N}$ is very large.  But if 
${\cal N}$ is large, it is difficult to solve Eqs.~(\ref{crev}) numerically.  
Indeed, the number of system  states (the dimension of Hilbert space) is 
\cite{Huang}
\begin{equation}
D = {(N + M - 1)! \over N!~(M-1)!}
\label{dim}
\end{equation}
if $N$ is the number of particles and $M$ the number of particle states 
they occupy.  Clearly, we will only be able to simulate systems with
relatively small $N$ and $M$. (The simulation described below has 
$M=5$ and $N=50$, hence $D$ = 316 251 and ${\cal N}$ = 10.)  Since 
${\cal N}$ cannot be very large in a numerical simulation, how then 
does one simulate a system in the condensed regime?

We may contrive a toy model in the condensed regime by setting 
$\Lambda_{kl}^{ij} = 0$  when $\Omega_{kl}^{ij} \neq 0$. In the 
example studied below, $M = 5$ and the states are equally spaced: 
$\omega_j = j \omega_1$ for $j$ = 1,2, .. 5. The $\Lambda_{kl}^{ij}$ 
all vanish except $\Lambda_{23}^{14}$, $\Lambda_{24}^{15}$, 
$\Lambda_{34}^{25}$ and their complex conjugates $\Lambda_{14}^{23}$, 
$\Lambda_{15}^{24}$ and $\Lambda_{25}^{34}$.  The toy Hamiltonian 
is therefore
\begin{eqnarray}
H &=& \omega_1 a_1^\dagger a_1 +  2 \omega_1 a_2^\dagger a_2
+ 3 \omega_1 a_3^\dagger a_3 + 4 \omega_1 a_4^\dagger a_4
+ 5 \omega_1 a_5^\dagger a_5
+ \Lambda_{23}^{14} a_2^\dagger a_3^\dagger a_1 a_4
+ \Lambda_{23}^{14*} a_1^\dagger a_4^\dagger a_2 a_3\nonumber\\
&+& \Lambda_{24}^{15} a_2^\dagger a_4^\dagger a_1 a_5
+ \Lambda_{24}^{15*} a_1^\dagger a_5^\dagger a_2 a_4
+ \Lambda_{34}^{25} a_3^\dagger a_4^\dagger a_2 a_5
+ \Lambda_{34}^{25*} a_2^\dagger a_5^\dagger a_3 a_4 \ .
\label{toyHs}
\end{eqnarray}
The toy model is a far cry from a description of the cold axion fluid 
but it is in the condensed regime in the sense that its evolution is 
governed by Eq.~(\ref{crev}), and that evolution is non trivial.  We 
may ask how quickly does the toy system reach thermal equilibrium 
starting from an arbitrary initial state $|{\cal N}_1, {\cal N}_2, 
{\cal N}_3, {\cal N}_4, {\cal N}_5>$.  The initial state determines 
the energy $E$ through Eq.~(\ref{eeig}).  The (microcanonical ensemble) 
thermal averages of the ${\cal N}_j$ are
\begin{equation}
\bar{\cal N}_j = 
{\displaystyle\sum_{\{{\cal N}\}} {\cal N}_j \delta(E - E(\{{\cal N}\})) 
\over \displaystyle\sum_{\{{\cal N}\}} \delta(E - E(\{{\cal N}\}))}~~\ .
\label{thav}
\end{equation}
The $\bar{\cal N}_j$ were calculated numerically.  They differ generally 
from the Bose-Einstein distribution 
\begin{equation}
\bar{\cal N}_{j,{\rm BE}} = {1 \over e^{(\omega_j - \mu)/T} - 1}~~~\ ,
\label{BEd}
\end{equation} 
where $T$ is temperature and $\mu$ is chemical potential, because
the toy model is not in the thermodynamic limit.  

We calculate the quantum mechanical averages 
\begin{equation}
<{\cal N}_j(t)> = \sum_{\{{\cal N}\}} {\cal N}_j |\Psi(\{{\cal N}\},t)|^2
\label{quav}
\end{equation}
as a function of time and ask how quickly do they reach the thermal 
averages $\bar{\cal N}_j$.  Fig. 1 shows the result of a particular 
simulation with $N=50$.  The initial state is $|20, 5, 15, 5, 5>$, 
hence $E = 120~\omega_1$. The thermal averages for that value of $E$ are
$\{\bar{\cal N}\} = (15.25, 14.72, 9.33, 6.16, 4.53)$.  The evolution was
computed using Eqs.~(\ref{crev}) and (\ref{quav}) with $\Lambda_{23}^{14}
= \Lambda_{24}^{15} = \Lambda_{34}^{25} = 0.1$.  The time step was 
$\Delta t = 10^{-6}$, sufficiently small that energy is conserved 
within 1\%.  Fig. 1 shows the  $<{\cal N}_j(t)>$ as a function of 
$t$.  Their values at $t$ = 2.3 are $(15.57, 14.61, 9.42, 6.29,
4.59)$.  There is no doubt that the system thermalizes. 
Eq.~(\ref{est}) estimates the thermalization rate to be 
$\Gamma \sim \sqrt{3} \times 10 \times 0.1 = 1.7$, i.e. the 
thermalization time $1/\Gamma$ is 0.6 .  That estimate is 
consistent with the time evolution shown in Fig. 1.  We ran 
simulations for a number of different initial conditions.  In 
each case, the ${\cal N}_j(t)$ approached the appropriate thermal 
averages on a time scale of order $1/\Gamma \sim$ 0.6. 

It may appear unsurprising (and anti-climactic) that the simulations 
agree with the thermalization rate that we expected.  Let us keep in 
mind, however, that there have been surprises in past simulations of 
thermalization.  Fermi, Pasta and Ulam \cite{FPU} performed a famous 
numerical experiment where a set of classical coupled oscillators is 
evolved in time according to a Hamiltonian of the form
\begin{equation}
H = \sum_{j=1}^M 
\left({p_j^2 \over 2m} + {1 \over 2} m \omega_j^2 x_j^2\right)
+ \sum_{ijkl} \Lambda_{ijkl}~x_i x_j x_k x_l~~~\ .
\label{classic}
\end{equation} 
They found that the system does not thermalize even after a very 
long time.  We repeated this relatively simple simulation with $M=10$ 
oscillators and also found that the system does not approach thermal 
equilibrium, even after a very long time.  It is reassuring that, in 
contrast, the system of bosonic oscillators behaves in the way we 
expect.  As far as we know, ours is the first simulation of the 
approach to thermal equilibrium of a set of coupled quantum 
oscillators. 

\section{Axion cosmology revisited}

In this section, we consider how the late thermalization of axions 
and other species modifies the standard cosmological model.  It is 
assumed that the estimates of the thermalization rates in Section III.D 
are correct.  We find that cold axions may briefly thermalize after they 
are first produced during the QCD phase transition as a result of their 
$\lambda \phi^4$ self-interactions.  However, this first thermalization 
era promptly ends and the axions are then decoupled until the photon 
temperature reaches approximately 500 eV.  At that time the axions 
thermalize through their gravitational self-interactions.  They form
 a BEC, meaning that most axions go to the lowest energy state available 
to them.   The axion correlation length grows to sizes of order the horizon.  
The ratio $\Gamma_a/H$, of the cold axion thermalization rate to the Hubble 
rate, grows then as $a(t)^{-3} t^3$, the axions thermalize on ever shorter 
time scales relative to the age of the universe and the axion state tracks 
the lowest energy state at all times.  We find that baryons enter into 
thermal contact with the cold axions soon after the axion correlation 
length has grown to be of order the horizon.   It appears possible, 
although by no means certain, that baryons, photons and axions all 
reach the same temperature at about the time of equality between 
matter and radiation.  In that case and assuming that the initial 
photon chemical potential is negligible, the photon temperature drops 
by the factor $({2 \over 3})^{1 \over 4}$ = 0.9036 compared to the 
standard cosmological model.  We derive the implications of this for 
observations of cosmological parameters, specifically the baryon to 
photon ratio $\eta_{\rm BBN}$ at the time of primordial nucleosynthesis 
and the effective number $N_{\rm eff}$ of neutrino thermal degrees 
of freedom at decoupling.  We consider whether neutrinos may also come 
into thermal contact with the cold axions, and what this would imply 
for $\eta_{\rm BBN}$ and $N_{\rm eff}$.  Finally we show that axion 
rethermalization by gravitational self-interactions is sufficiently fast 
that axions that are about to fall into a galactic gravitational potential 
well share their angular momenta.  By almost all going to the lowest energy 
state for given total angular momentum, the axions acquire net overall rotation, 
implying $\vec{\nabla} \times \vec{v} \neq 0$ where $\vec{v}(\vec{r})$ is the 
velocity field of the infalling axions.  In contrast, the velocity field of 
WIMP dark matter is irrotational.  This provides a means to distinguish 
axions from  WIMPs by observing the inner caustics of galactic halos.
As mentioned in the Introduction, the evidence for caustic rings is 
consistent with axions and inconsistent with WIMPs.  

\subsection{QCD epoch}

Cold axions are produced when the axion mass turns on during the 
QCD phase transition.  The critical time is $t_1$ defined by 
$m(t_1) t_1 = 1$ where $m(t)$ is the axion mass. At temperatures
well above 1 GeV, $m \simeq 0$ whereas at temperatures well below 
100 MeV, $m$ has its zero temperature value. A standard calculation 
yields \cite{axdm}
\begin{equation}
t_1 \simeq 2 \cdot 10^{-7}~{\rm sec} 
\left({f_a \over 10^{12} {\rm GeV}}\right)^{1 \over 3}~~,~~
T_1 \simeq 1~{\rm GeV} 
\left({10^{12} {\rm GeV} \over f_a}\right)^{1 \over 6}
\label{t1}
\end{equation}
where $T_1$ is the photon temperature at $t_1$, and $f_a$ is 
the axion decay constant.  How many cold axions are produced 
depends in part on whether inflation occurs before or after 
the Peccei-Quinn phase transition \cite{axcos}.  If before, 
there are contributions to the cold axion density from vacuum 
realignment, string decay and wall decay.  The size of the 
string decay contribution has been the matter of debate and 
controversy.  If after, the contributions from string and wall 
decay are absent and the contribution from vacuum realignment 
may be accidentally suppressed if the initial value of the 
homogenized axion field happens to lie close to the CP conserving 
minimum.  Allowing for these and other uncertainties, we write the 
cold axion number density at time $t_1$ as 
\begin{equation}
n(t_1) = X {f_a^2 \over t_1}~~\ .
\label{X}
\end{equation}
$X$ is of order two if there is no inflation after the PQ 
phase transition and the contribution from string decay is 
of the same order as that from vacuum realignment, $X$ is 
of order ten if there is no inflation after the PQ phase 
transition and the contribution from string decay dominates 
over that from vacuum realignment(the contribution from wall 
decay is thought to be subdominant always), and $X$ is of 
order one half the square of the sine of the misalignment 
angle if there is inflation after the PQ phase transition.  
The misalignment angle is a random number, between 0 and 
$2\pi$, which has the same value in our whole visible 
universe in this case.
 
After $t_1$ the number of axions is conserved in the sense that 
all axion number changing processes, such as axion decay to two 
photons, happen on time scales vastly longer than the age of the
universe.  Eqs.~(\ref{t1}) and (\ref{X}) imply then the cold axion 
number density as a function of time
\begin{equation}
n(t) \simeq {4 \cdot 10^{47} \over {\rm cm}^3} X
\left({f_a \over 10^{12} {\rm GeV}}\right)^{5 \over 3}
\left({a(t_1) \over a(t)}\right)^3~~~\ .
\label{nat}
\end{equation} 
Most of the cold axions are non-relativistic shortly after $t_1$
because the axion momenta are of order ${1 \over t_1}$ at time 
$t_1$ and vary with time as $a(t)^{-1}$.  Hence we may obtain 
the axion energy density today \cite{axdm,axcos}
\begin{equation}
\Omega_a \equiv  {8 \pi G \over 3 H_0^2} m n(t_0)
\simeq 0.3~X \left({f_a \over 10^{12} {\rm GeV}}\right)^{7 \over 6}~~\ .
\end{equation}
The velocity dispersion of cold axions is 
\begin{equation}
\delta v (t) \sim {1 \over m t_1} {a(t_1) \over a(t)}
\label{veldis}
\end{equation}
if each axion remains in whatever state it is in, i.e. if axion
interactions are negligible.  We refer to this case as the limit 
of decoupled cold axions (see subsection II.C.1).  If decoupled, 
the average state occupation number of the cold axions is \cite{CABEC}
\begin{equation}
{\cal N} \sim n {(2 \pi)^3 \over {4 \pi \over 3} (m \delta v)^3}
\sim 10^{61}~X \left({f_a \over 10^{12} {\rm GeV}}\right)^{8 \over 3}~~\ .
\label{calN}
\end{equation}
Because ${\cal N}$ is huge, we may expect cold axions to be a 
Bose-Einstein condensate.  Their energy dispersion $\delta \omega = 
{1 \over 2} m (\delta v)^2$ is much smaller than the critical 
temperature
\begin{equation}
T_c(t) = \left({\pi^2 n(t) \over \zeta(3)}\right)^{1 \over 3}
\simeq 300~{\rm GeV}~X^{1 \over 3}~ 
\left({f_a \over 10^{12} {\rm GeV}}\right)^{5 \over 9}
{a(t_1) \over a(t)}
\label{Tc}
\end{equation}
for BEC.  Note that Eq.~(\ref{Tc}) gives the critical temperature
for a {\it relativistic} ($T_c >> m$) Bose gas. 

Actually, there are four conditions for a gas to form a Bose-Einstein 
condensate: 1) the particles are identical bosons, 2) their number is 
conserved, 3) their phase space density in units of $h^{-3} = 
(2 \pi \hbar)^{-3}$ is of order one or larger (equivalently their 
average quantum state occupation number ${\cal N}$ is of order one or 
larger), and 4) the particles are in thermal equilibrium.  The first 
three conditions are clearly satisfied in the case of cold axions.  
The fourth condition is the critical one since it is not obviously 
satisfied.  Indeed, axions are thought to be very weakly interacting.

Axions are in thermal equilibrium if their relaxation rate $\Gamma_a$ is 
large compared to the Hubble expansion rate $H$.  We noted in Section III 
that the formula for the relaxation rate differs whether the particles in 
the fluid are in the `particle kinetic' regime or in the `condensed' regime. 
At time $t_1$, cold axions are at the borderline between the two regimes since
\begin{equation}
\delta \omega (t_1) = {1 \over 2} m(t_1) (\delta v (t_1))^2
\sim {1 \over 2 m(t_1) t_1^2} = H(t_1)~~\ .
\label{border2}
\end{equation}
To see whether the axions thermalize by $\lambda \phi^4$ self-interactions, 
we use Eqs.~(\ref{xsec}) and (\ref{relax2}) in the particle kinetic regime, 
and Eq.~(\ref{relcon}) in the condensed regime.  At $t_1$ both estimates 
give the same answer\cite{CABEC}, namely:
\begin{equation} 
\Gamma_\lambda(t_1) \sim H(t_1)
\label{lamt1}
\end{equation}
indicating that the axions barely thermalize at time $t_1$ by 
$\lambda \phi^4$ self-interactions.  After $t_1$ we must use 
Eq.~(\ref{relcon}) since we are in the condensed regime then. 
It informs us that 
$\Gamma_\lambda (t) / H(t) \propto a(t)^{-3} t \propto t^{-{1 \over 2}}$, 
i.e. that even if axions thermalize at time $t_1$ they stop doing so 
shortly thereafter.  Nothing much changes as a result of this brief
epoch of thermalization since in either case, whether it occurs or 
not, the correlation length $\ell(t) \sim t_1 a(t)/a(t_1)$.

To see whether gravitational interactions cause the cold axions 
to form a BEC, we use Eq.~(\ref{relcong}).  It implies
\begin{equation}
\Gamma_g(t) / H(t) \sim 8 \pi G n m^2 \ell^2 t
\sim 5 \cdot 10^{-7}~{a(t_1) \over a(t)} {t \over t_1} X 
\left({f_a \over 10^{12} {\rm GeV}}\right)^{2 \over 3}
\label{GgoH}
\end{equation}
once the axion mass has reached its zero temperature value, 
shortly after $t_1$.  Gravitational self-interactions are 
too slow to cause thermalization of cold axions near the 
QCD phase transition but, because 
$\Gamma_g / H \propto a^{-1}(t) t \propto a(t)$, they do 
cause the cold axions to thermalize later on \cite{CABEC}.

\subsection{Axion BEC formation}

The RHS of Eq.~(\ref{GgoH}) reaches one at a time $t_{\rm BEC}$ 
when the photon temperature is of order 
\begin{equation}
T_{\rm BEC} \sim 500~{\rm eV}~X 
\left({f_a \over 10^{12}~{\rm GeV}}\right)^{1 \over 2}~~\ .
\label{Tbec}
\end{equation}
The axions thermalize then and form a BEC as a result of their 
gravitational self-interactions.  The whole idea may seem far-fetched 
because we are used to think that gravitational interactions among 
particles are negligible.  The axion case is special, however, because 
almost all particles are in a small number of states with very long 
de Broglie wavelength, and gravity is long range.  By gravitational 
self-interactions the axions modify their momentum distribution till 
their entropy is maximized for the available energy, which in this 
case means that they form a BEC.

Axion BEC causes the correlation length to increase.  Indeed 
in an infinite volume, when all particles are in the lowest 
energy state, the momentum dispersion is theoretically zero 
and the correlation length infinite.  This ideal state never 
occurs because thermalization and hence BEC formation 
are constrained by causality.  The axions in one horizon 
are unaware of the doings of axions in the next horizon.  
Hence we expect the correlation length $\ell$, which may 
now be thought of as the size of condensate patches, to 
become of order but less than the horizon.  The growth in 
the correlation length causes the thermalization to accelerate; 
see Eq.~(\ref{GgoH}).  Once $l$ is some fraction of $t$, 
$\Gamma_g(t)/H(t) \propto a(t)^{-3}t^3$, implying that 
thermalization occurs on ever shorter time scales compared 
to the Hubble time.

\subsection{Thermal contact with other species}

The rate at which non-relativistic species such as baryons change their 
momentum distribution through gravitational interactions with the cold 
axion fuid is given by Eq.~(\ref{relb}).  The momentum dispersion of 
baryons is of order  $\Delta p \sim \sqrt{3 m_B T}$ where $T$ is the 
photon temperature.  We will assume here that cold axions are the 
bulk of the dark matter.  The Friedmann equation implies then
\begin{equation}
4 \pi G n m \sim {3 \over 8 t^2} 
\left({t \over t_{\rm eq}}\right)^{1 \over 2}
\label{Fried}
\end{equation}
for $t < t_{\rm eq}$, where $t_{\rm eq}$ is the time of equality 
between matter and radiation. Hence
\begin{equation}
\Gamma_B/H \sim {1 \over 4} {\ell \over \sqrt{t t_{\rm eq}}} 
\sqrt{3 m_B \over T} 
\sim ~{\ell \over t} \left({t \over t_{\rm BEC}}\right)^{3 \over 4}
\left({500~{\rm eV} \over T_{\rm BEC}}\right)^{3 \over 2}~~~\ .
\label{gboh}
\end{equation}
Eq.~(\ref{gboh}) shows that baryons reach thermal contact with 
the axion BEC when $\ell$ becomes of order $t$ or soon after that.  

Photons are in thermal contact with the baryons, but the nature and 
degree of this thermal contact are changing at the time of axion BEC 
formation \cite{Hu}.  Baryons interact with electrons by Coulomb 
scattering.  Electrons interact with photons by Compton scattering, 
double Compton scattering and bremsstrahlung.  Above approximately 1 
keV photon temperature, double Compton scattering and bremsstrahlung 
assure chemical equilibrium between baryons and photons (the number 
of photons is not conserved in these processes). Below approximately 
1 keV photon temperature, Compton scattering is the only important 
interaction remaining.  It maintains kinetic, but not chemical, 
equilibrium between baryons and photons till approximately 100 eV 
photon temperature.  Below 100 eV, the degree of kinetic equilibrium 
progressively diminishes till approximately 2 eV, when it disappears
altogether.  

In any case, as long as there is only thermal contact between baryons 
and a few low momentum modes of the axion field, only a very small 
amount of energy can be exchanged between the axion field and the other 
species.  However, as time goes on, higher and higher momentum modes 
of the axion field reach thermal contact with its highly occupied low 
momentum modes.  Eq.~(\ref{relr}) gives the relaxation rate $\Gamma_r$ 
of the axion field as a whole, including relativistic states.  The 
relaxation rate of photons $\Gamma_\gamma$, Eq.~(\ref{relg}), is of 
the same order of magnitude.   Using Eq.~(\ref{Fried}), we have
\begin{equation}
\Gamma_\gamma/H \sim \Gamma_r/H \sim {3 \ell \over 4 t} 
\left({t \over t_{\rm eq}}\right)^{1 \over 2}
\label{crit}
\end{equation}
for $t < t_{\rm eq}$.  After equality ($t > t_{\rm eq}$), 
$\Gamma_r/H \propto n \ell t \propto a(t)^{-3} t^2$ 
= constant.  Hence, if $\Gamma_r/H$ does not reach one 
before equality, it will remain less than one forever afterwards.  

Hence we consider at this stage two possibilities, which we call 
cases A and B.  In case A,  $\Gamma_r/H$ does not reach one 
before equality, (perhaps because $\ell$, although proportional to 
$t$ is much less than $t$, e.g. $\ell = t$/100), and hence thermal 
contact gets established only between baryons and low momentum modes 
of the axion field.  In case B, $\Gamma_r/H$ does reach one before 
equality and thermal equilibrium is reached between baryons, axions 
and photons.  This equilibrium is kinetic only since gravitational 
interactions conserve particle number for all the species involved.

We should ask whether neutrinos may also reach thermal contact 
with the highly occupied low momentum axion modes, in which case 
neutrinos, axions, baryons and photons would all reach the same 
temperature.  We believe this possibility, which we call case C, 
unlikely for the following reason.  Eq.~(\ref{general}) does not 
apply to degenerate fermions because of Pauli blocking.  Cosmic 
neutrinos are semi-degenerate since they have a thermal distribution 
with zero chemical potential.  Because of partial Pauli blocking, 
their thermalization is slower than that of relativistic axions, 
Eq.~(\ref{relr}).  Since relativistic axions only barely reach 
thermal contact with the highly occupied low momentum modes of 
the axion field if they do so at all, and thermal contact between 
those low momentum modes of the axion field and neutrinos is delayed 
relative to relativistic axions, it appears most likely that neutrinos 
remain decoupled from the axions at all times.  Although we believe 
case C unlikely, we consider its implications below along with 
the other two cases.

\subsection{Implications for cosmological parameters}

In case A, the baryons reach the same temperature as the 
low momentum states of the axion field but only a very small 
amount of energy is transferred from baryons to axions, because 
the relativistic states of the axion field remain decoupled from 
its low momentum states.  The cosmological parameters are the same 
as in the standard cosmological model with WIMP dark matter.

In case B, baryons, axions and photons all reach the same temperature.  
The axions are heated whereas the photons are cooled.  The number of 
photons is unchanged implying that their chemical potential stays at
zero or, if photons have a small negative chemical potential to start 
with, increases till it reaches zero.  There is photon Bose-Einstein 
condensation, i.e. the excess photons condense into the lowest energy 
available state, which is a plasma wave with vanishing wavevector.  With 
zero chemical potential, the photon energy spectrum is Planckian, as 
required by observation.  Energy conservation implies
\begin{equation}
\rho_{\gamma i} = {\pi^2 \over 15} T_{\gamma i}^4 
=  \rho_{\gamma f} + \rho_{a f} = 
{\pi^2 \over 30} T_{\gamma _f}^4 (2 + 1)~~\ .
\label{enconB}
\end{equation}
Hence $T_{\gamma f} = 
\left({2 \over 3}\right)^{1 \over 4} T_{\gamma i}
= 0.9036~T_{\gamma i}$ in case B.  In Eq.~(\ref{enconB})
we neglected the contributions from the initial state axions, 
the baryons, and the condensed photons in the final state. 
This is justified because these contributions are respectively 
of order $10^{-22}$, $10^{-9}$ and $10^{-9}$ relative to the 
contributions that are included in Eq.~(\ref{enconB}).  We also 
assumed that the initial chemical potential of the photons is 
zero.  Finally, we ignored the fact that, while the transfer of 
energy between photons and axions takes place, the expansion of 
the universe affects photons and axions differently because 
axions have mass.  The corresponding error is of order 
$m/T_{\rm eq} \sim 10^{-5}$, or less. 

After the axions have been heated by thermal contact with 
the photons, the fraction of axions in thermally excited 
states is of order
\begin{equation}
\left({T_{\rm eq} \over T_c(t_{\rm eq})}\right)^3 \sim 
4 \cdot 10^{-8}~{1 \over X}~
\left({10^{12}~{\rm GeV} \over f_a}\right)^{13 \over 6}~~\ .
\label{fraceq}
\end{equation} 
All others are condensed in the lowest energy available state.
The condensed axions are cold dark matter, with the special 
properties that are the topic of this paper. The axions in 
thermally excited states contribute one bosonic degree of 
freedom to radiation.  After axions, baryons and photons 
have all reached the same temperature, further thermalization
is suppressed because, once most of the axions are in the 
same state, the only interactions that are enhanced by large 
occupation numbers are interactions without momentum transfer
(see Section III.D).

In case B, cosmological parameters are modified compared to their 
values in the standard model with WIMP dark matter. Since the photons 
cool between the epoch of primordial nucleosynthesis (BBN, for short) 
and decoupling, and their temperature at decoupling is known, their 
temperature at BBN is larger by the factor 
$\left({3 \over 2}\right)^{1 \over 4}$ compared to the standard model.  
Because baryon number is conserved, the baryon to photon ratio 
$\eta \equiv n_b/n_\gamma$is smaller at BBN than at decoupling.  The 
baryon to photon ratio at decoupling is reliably determined from 
measurements of the anisotropy spectrum of the cosmic microwave 
radiation.  The latest WMAP result \cite{WMAP} is 
$\eta_{\rm dec} = (6.190 \pm 0.145)~10^{-10}$.  Since the photon 
number density $n_\gamma \propto T^3$, the value of $\eta$
at BBN is 
\begin{equation}
\eta_{\rm BBN} = \left({2 \over 3}\right)^{3 \over 4} \eta_{\rm dec}
= (4.57 \pm 0.11)~10^{-10}
\label{eta}
\end{equation}
in case B, whereas $\eta_{\rm BBN} = \eta_{\rm dec}$ in the standard 
model.  Under the assumption that there are three neutrino species, 
$\eta_{\rm BBN}$ is the main parameter controlling the primordial 
abundances of the light elements.  In the standard model, there is 
generally good agreement between the observed and predicted abundances 
of three light elements (D, $^4$He, $^3$He) but there is a discrepancy
for $^7$Li, the so-called `Lithium problem' \cite{Lith}.  The Lithium 
problem is alleviated in case B, and perhaps solved altogether, because 
of the lower value of $\eta_{\rm BBN}$.  On the other hand, the agreement 
that occurs in the standard model between the observed and predicted 
abundance of deuterium is spoiled \cite{Li7}.

Case B differs also from the standard cosmological model in that it 
has more radiation in collisionless species (axions and neutrinos).  
The radiation content of the universe is commonly given in terms of 
the effective number $N_{\rm eff}$ of thermally excited neutrino 
degrees of freedom, defined by 
\begin{equation}
\rho_{\rm rad} = \rho_\gamma \left[ 1 + N_{\rm eff}~{7 \over 8}
\left({4 \over 11}\right)^{4 \over 3} \right]
\label{neff}
\end{equation}
where $\rho_{\rm rad}$ is the total energy density in radiation 
(including photons, neutrinos and axions) and $\rho_\gamma$ is 
the energy density in photons only.  The standard model predicts
$N_{\rm eff}$ = 3.046, slightly larger than 3 because the three
neutrinos heat up a little during $e^+ e^-$ annihilation. In 
case B
\begin{equation}
\rho_{\rm rad} = \rho_\gamma + \rho_a + \rho_\nu = 
\rho_\gamma \left[ 1 + {1 \over 2} + 
(3.046) {7 \over 8} \left({4 \over 11}\right)^{4 \over 3} {3 \over 2}\right]
\label{rhoB}
\end{equation}
taking account of the fact that not only is there an extra species 
of radiation (axions) but also the contribution of neutrinos is boosted
because the photons have cooled relative to them.  Eq.~(\ref{rhoB}) 
implies $N_{\rm eff}$ = 6.77~.  At present, most measurements place
$N_{\rm eff}$ between 4 and 5, with one $\sigma$ uncertainties of 
order 1 \cite{WMAP,Neff}.  The tendency for the measured values to be 
larger than 3.046 has been taken sufficiently seriously to prompt 
proposals for new physics involving extra neutrino species or a 
neutrino asymmetry \cite{exneu}.  The Planck mission is expected 
to measure $N_{\rm eff}$ with much greater precision \cite{Ichi}.  
In so doing, it may shed light on the nature of dark matter.

In case C, the cosmological parameters differ from their standard 
values even more than in case B.  Energy conservation implies 
in case C
\begin{eqnarray}
\rho_{\gamma i} + \rho_{\nu i} &=&
{\pi^2 \over 30} T_{\gamma i}^4 \left[ 2 + 
(3.046) \cdot 2 \cdot {7 \over 8} 
\left({4 \over 11}\right)^{4 \over 3}\right]
\nonumber\\
&=& \rho_{\gamma f} + \rho_{a f} + \rho_{\nu f} 
= {\pi^2 \over 30} T_f^4 (2 + 1) + \rho_{\nu f}(T_f, \mu_{\nu f})
\label{enconC}
\end{eqnarray}
where $T_f$ is the common final temperature of photons, axions and 
neutrinos, and $\mu_{\nu f}$ is the final chemical potential of the 
neutrinos.  In writing Eq.~(\ref{enconC}) we make the same approximations 
as for Eq.~(\ref{enconB}).  When gravitational interactions establish 
kinetic equilibrium between axions, photons and neutrinos, the photons 
cool whereas the axions and neutrinos heat up.  Since the number of 
neutrinos is conserved in this process, the neutrinos acquire a 
negative chemical potential.  To obtain $T_f$ we solve numerically 
\begin{equation}
n_{\nu i} = {3 \zeta(3) \over 4 \pi^2} T_{\nu i}^3 = n_{\nu f} =
\int {d^3 k \over (2 \pi)^3} 
{1 \over e^{(k - \mu_{\nu f}) \over T_f} + 1}
\label{nnui}
\end{equation}
with $T_{\nu i} = \left({4 \over 11}\right)^{1 \over 3} T_{\gamma i}$, 
and Eq.~(\ref{enconC}) with 
\begin{equation}
\rho_{\nu f} = 2 (3.046) \int {d^3 k \over (2 \pi)^3}
{k \over e^{(k - \mu_{\nu f}) \over T_f} + 1}~~~\ . 
\label{rhonuf}
\end{equation}
This yields $T_f = 0.873 T_{\gamma i}$ and $\mu_{\nu f} = - 0.65 T_f$.  
The photons cool even more than in case B ($T_f = 0.904 T_{\gamma i}$)
because their heat is transferred to neutrinos as well as axions. In 
case C, the baryon to photon ratio at the epoch or primordial 
nucleosynthesis is 
\begin{equation}
\eta_{\rm BBN} = (0.873)^3 \eta_{\rm dec} = 
(4.12 \pm 0.10) 10^{-10}
\label{etaC}
\end{equation}
using again the WMAP value $\eta_{\rm dec} = (6.190 \pm 0.145) 10^{-10}$.  
The total radiation density is also higher. Equating the RHS of 
Eq.~(\ref{enconC}) with the RHS of Eq.~(\ref{neff}) yields 
$N_{\rm eff} = 8.3$.
  
\subsection{Tidal torquing with axion BEC}

Let us consider a region of size $L$ inside of which the axion state 
(i.e. the state that most axions are in) stops being the lowest energy 
available state because the background is time dependent.  Under what 
conditions is thermalization by gravitational self-interactions sufficiently 
fast that the condensed axions remain in the lowest energy state as the 
background evolves?  We use the same heuristic reasoning that led us to 
Eq.~(\ref{general}), which we later verified by more formal derivations 
in a number of cases in Section III.E.  We expect that the axion BEC 
rethermalizes provided the gravitational forces produced by the BEC 
are larger than the typical rate $\dot{p}$ of change of axion momenta 
required for the axions to remain in the lowest energy state.  The 
gravitational forces are of order $4 \pi G n m^2 \ell$.  In this case, 
the correlation length $\ell$ must be taken to be of order the size $L$ 
of the region of interest since the gravitational fields due to axion BEC 
outside the region do not help the thermalization of the axions within the 
region.  Hence the condition is
\begin{equation}
4 \pi G n m^2 L \gtrsim \dot{p}~~~\ .
\label{thercon}
\end{equation}
We now apply this criterion to the question whether axions rethermalize 
sufficiently quickly that they share angular momentum when they are about 
to fall into a galactic gravitational potential well.

We use the self-similar infall model of galactic halo formation 
to estimate $L$ and $\dot{p}$.  $L$ is of order a few times the 
turnaround radius $R(t)$, say $L(t) \sim 3 R(t)$, whereas 
$p(t) \sim m v_{\rm rot}(t) j_{\rm max}$ where $v_{\rm rot}$ 
is the rotation velocity and $j_{\rm max}$ is the dimensionless
number characterizing the amount of angular momentum of the halo.
In the self-similar model, $v_{\rm rot}(t) \sim R(t)/t$ and 
$R(t) \propto t^{{2 \over 3} + {2 \over 9 \epsilon}}$ where 
$\epsilon$ is in the range 0.25 to 0.35 \cite{STW}.  Assuming 
that most of the dark matter is axions, the Friedmann equation 
implies
\begin{equation}
4 \pi G n m = {3 \over 2} H(t)^2 = {2 \over 3 t^2}
\label{Fried2}
\end{equation} 
for $t > t_{\rm eq}$.  The LHS of Eq.~(\ref{thercon}) is 
therefore of order
\begin{equation}
2 m {R(t) \over t^2} \sim 2 m v_{\rm rot}(t) {1 \over t}
\label{LHS}
\end{equation}
whereas its RHS is of order 
\begin{equation}
{d \over dt} \left[ m j_{\rm max} v_{\rm rot}(t_0) 
\left({t \over t_0}\right)^{- {1 \over 3} + {2 \over 9 \epsilon}} \right]
= m j_{\rm max} v_{\rm rot}(t){1 \over t} 
({2 \over 9 \epsilon} - {1 \over 3})~~~\ .
\label{RHS}
\end{equation}
The typical value of $j_{\rm max}$ is 0.18 .  Hence Eq.~(\ref{thercon})
is satisfied at all times from equality till today by a margin of 
order ${2 \over j_{\rm max} ({2 \over 9 \epsilon} - {1 \over 3})}
\sim 30$ .

We conclude that the axion BEC does rethermalize before falling 
into the gravitational potential well of a galaxy.  Most axions
go to the lowest energy state consistent with the total angular 
momentum acquired from neighboring inhomogeneities through tidal 
torquing \cite{TTT}.  That state is a state of rigid rotation on 
the turnaround sphere, implying $\vec{\nabla} \times \vec{v} \neq 0$ 
where $\vec{v}$ is the velocity field of the infalling axions.  In 
contrast, the velocity field of WIMP dark matter is irrotational.  
The inner caustics of galactic halos are different in the two cases.  
Axions produce caustic rings whereas WIMPs produce the `tent-like' 
caustics described in ref.~\cite{inner}.  There is evidence for the 
existence of caustic rings in various galaxies at the radii predicted 
by the self-similar infall model.  For a review of this evidence 
see ref. \cite{MWhalo}.  It is shown in ref. \cite{case} that the 
phase space structure of galactic halos implied by the evidence 
for caustic rings is precisely and in all respects that predicted 
by the assumption that the dark matter is a rethermalizing BEC.

\section{Summary}

The purpose of this paper was to investigate when and to what extent 
cold axions differ from ordinary cold dark matter, such as WIMPs and 
sterile neutrinos.  

In Section II, we showed that cold axions behave as ordinary cold 
dark matter on all scales of observational interest when they are 
decoupled.  Observable differences between cold axions and ordinary 
cold dark matter occur only when the axions self-interact.  When 
the axions self-interact at a sufficiently high rate, they thermalize 
and form a Bose-Einstein condensate.  Before that time, the axions 
are described by a free classical field and are indistinguishable from 
ordinary cold dark matter on all scales of observational interest.  After 
Bose-Einstein condensation, almost all axions are in the same state.  
If that state is time independent, axions again behave as ordinary 
cold dark matter on all scales of observational interest.  Observable 
differences occur if and only if the axions rethermalize so that the 
axion state tracks the lowest energy state.

In Section III, we calculated the thermalization rates of cold axions 
through $\lambda \phi^4$-type self-interactions and through gravitational 
self-interactions.  We described the axion field as a set of coupled 
quantum-mechanical oscillators and asked how quickly such a set of 
oscillators approaches thermal equilibrium.  We found that there 
are two distinct regimes: the `particle kinetic regime' defined by 
the condition that the energy dispersion is large compared to the 
interaction rate ($\delta \omega >> \Gamma$), and the opposite 
`condensed regime' where $\delta \omega << \Gamma$.  We derived
the Boltzmann equation as an operator statement in the particle
kinetic regime.  However, cold axion thermalization occurs almost 
entirely in the condensed regime.  We derived estimates for the 
thermalization rate in the condensed regime, and applied the estimates
to cold axions and to other species (baryons, photons and relativistic 
axions) that may come into thermal contact with the cold axions.

In Section IV, we performed numerical simulations to check our 
estimates of thermalization rates in the condensed regime.  We 
constructed a toy model with 50 quanta distributed among 5 oscillators.  
The exact quantum-mechanical evolution of the toy model was solved 
numerically for a variety of initial conditions.  It was found that 
the toy system thermalizes and does so on the time scale predicted 
by the analytical estimate in Section III.

In Section V, we investigated what changes the thermalization of cold 
axions brings to axion cosmology.  It was found that cold axions 
thermalize by gravitational self-interactions when the photon 
temperature is of order 500 eV.  When they thermalize, the 
cold axions form a Bose-Einstein condensate.  The axion correlation 
length grows to a size of order but smaller than the horizon.  
Shortly thereafter, baryons come into thermal contact with the 
axions.  As time goes on, increasingly higher momentum modes of 
both the axion field and the photon field come into thermal 
contact with the cold axions.  It is possible, but by no means
certain, that full thermal equilibrium is established among 
axions, photons and baryons.  This would imply a drop in the photon 
temperature by the factor $\left({2 \over 3}\right)^{1 \over 4}$
= 0.9036, at about the time of equality between matter and radiation.
We derived the implications of this for cosmological parameters, 
specifically the baryon to photon ratio at the time of primordial 
nucleosynthesis and the effective number of neutrinos (a measure
of the radiation content of the universe) at the time of decoupling.
Finally we showed that cold axions thermalize sufficiently fast by 
gravitational self-interactions that the axions about to fall into 
a galactic gravitational well acquire a state of net overall rotation.  
In contrast, ordinary cold dark matter falls in with an irrotational 
velocity field.  The inner caustics of the galactic halo are different 
in the two cases.  The occurrence of caustic rings of dark matter in 
galactic halos is inconsistent with ordinary cold dark matter, but 
consistent with axion BEC.

\begin{acknowledgments}
We have benefitted from conversations with many colleagues.  In 
particular, we would like to thank Georg Raffelt, Lawrence Widrow, 
Edward Witten, Jorge Gamboa, Francisco Mendez, Sergei Obukhov, 
David Lyth, Charles Thorn, Charles Sommerfield, Richard Woodard,
Leslie Rosenberg and Gary Steigman.  We are grateful to the UF HPC 
Center for computer time.  P.S. would like to thank the CERN Theory 
Group and the organizers of the DMUH11 workshop for their support and 
hospitality while working on this paper.  This work was supported in 
part by the U.S. Department of Energy under grant DE-FG02-97ER41209. 
 
\end{acknowledgments}


\newpage



\newpage

\begin{figure}
\begin{center}
\includegraphics[angle=270, width=150mm]{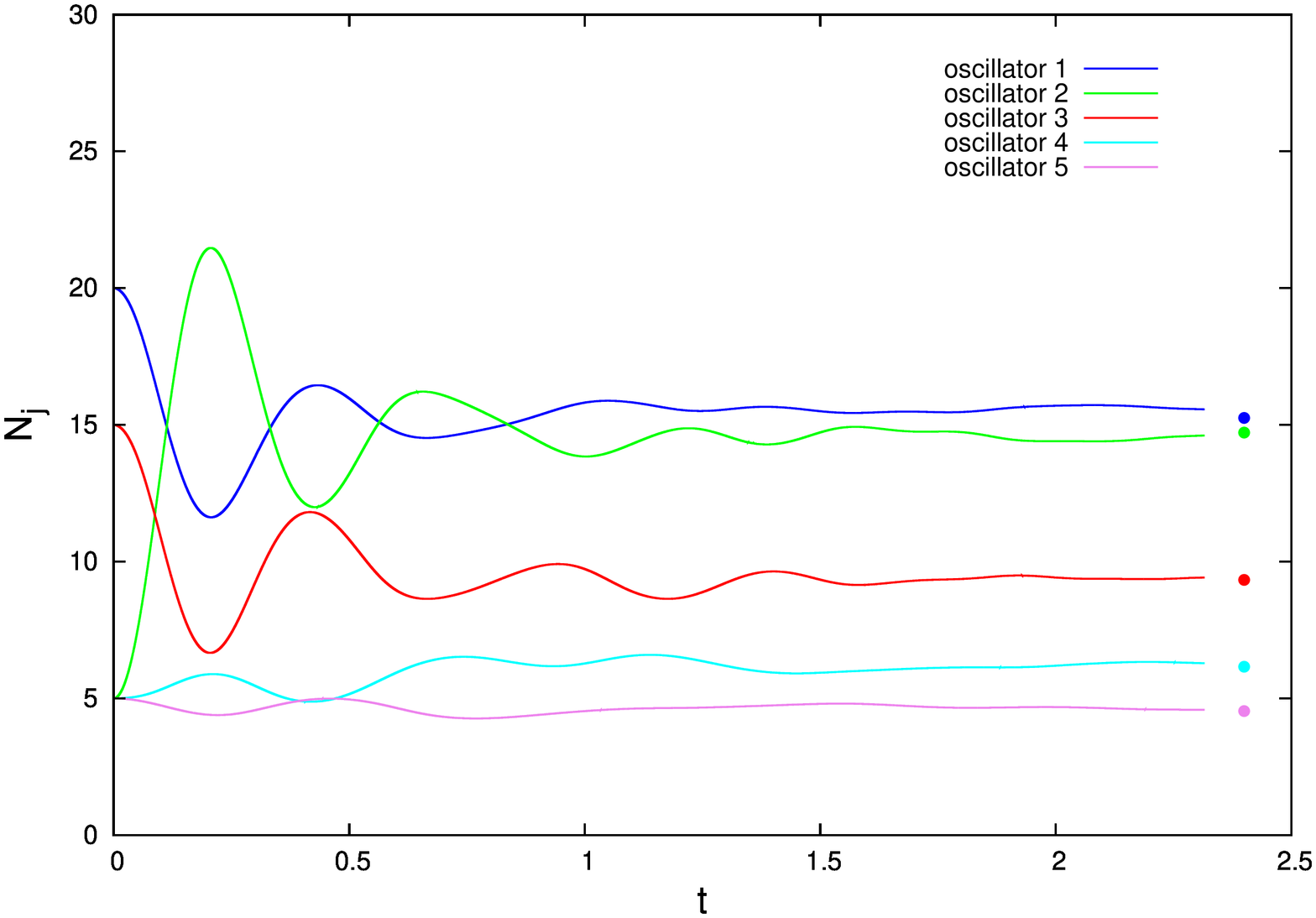}
\caption{Time evolution of the quantum mechanical average
particle number $<{\cal N}_j>, j = 1 .. 5$, in each of the 
five states of the toy model described in the text, starting 
with the initial state $({\cal N}_1, ... {\cal N}_5)$ = 
(20, 5, 15, 5, 5).  The dots indicate the thermal averages
$\bar{\cal N}_j$ for the corresponding total energy.  The 
system reaches thermal equilibrium on a time scale 
$1/\Gamma \sim 0.6$.}
\end{center}
\label{evol}
\end{figure}


\end{document}